\documentclass[12pt]{article}

\usepackage{epsfig,latexsym,amsfonts,amsmath,amsthm,amssymb,amsbsy,multirow,slashed,wasysym,textcomp,dsfont,comment,mathtools,cancel,cite,diagbox,datetime,appendix,BOONDOX-calo}
\usepackage{tocloft}
\usepackage{bbold}
\usepackage{sidecap}
\usepackage{adjustbox}
\usepackage{pgfplots}
\usepackage{oplotsymbl}
\usepackage{tikzpagenodes}
\usepackage{adforn}
\usepackage{tikz}
\usetikzlibrary{backgrounds}
\usetikzlibrary{patterns}
\usetikzlibrary{arrows.meta}
\usetikzlibrary{shapes}
\tikzstyle{bag} = [align=center]
\usetikzlibrary{decorations.pathmorphing}
\usepackage[normalem]{ulem}
\usepackage[mathscr]{eucal}
\usepackage[version=3]{mhchem}

\usepackage{hyperref}
\usepackage{subcaption}

 \newcommand{\badat}{\begin{alignedat}}
 \newcommand{\eadat}{\end{alignedat}}
 
 \def\be{\begin{equation}}
\def\ee{\end{equation}}

\def\p{\partial}

\usepackage{color}

\newcommand{\pink}[1]{\textcolor{\pink}{#1}}

\definecolor{dblue}{rgb}{0.2,0.50,0.80}

\def\bz{{\bar z}}

\def\bz{{\bar z}}

\numberwithin{equation}{section} 
\pgfplotsset{compat=1.17} 
\begin{document}

 \begin{titlepage}
  \thispagestyle{empty}
  \begin{flushright}
  \end{flushright}
  \bigskip

  \begin{center}

                  \baselineskip=13pt {\LARGE \scshape{
                  Celestial Holography}
         }

      \vskip1cm

   \centerline{   {Sabrina Pasterski},${}^\diamondsuit{}$
   {Monica Pate},${}^{\star}{}$
   {and Ana-Maria Raclariu}${}^\ddagger{}$
   }

\bigskip\bigskip
 \bigskip\bigskip

 {\em${}^\diamondsuit$ Princeton Center for Theoretical Science, Princeton, NJ 08544, USA}
 \vspace{.5em}

 \centerline{\em${}^\star$  Center for the Fundamental Laws of Nature, Society of Fellows, \& Black Hole Initiative, }
 \vspace{.25em}
 \centerline{\em Harvard University, Cambridge, MA 02138, USA}
 \vspace{.5em}

\centerline{\em${}^\ddagger$   Perimeter Institute for Theoretical Physics, Waterloo, ON N2L 2Y5, Canada
}

\bigskip\bigskip

\end{center}

\begin{abstract}
  \noindent

Celestial holography proposes a duality between the gravitational $\mathcal{S}$-matrix and correlators in a conformal field theory living on the celestial sphere. In this white paper, solicited for the 2022 Snowmass process, we review the motivation from asymptotic symmetries, fundamentals of the proposed holographic dictionary, potential applications to experiment and theory, and some important open questions.

\end{abstract}

\end{titlepage}

\setcounter{tocdepth}{2}

\tableofcontents
\vspace{3em}

\section{Motivation}

The formulation of a complete theory of quantum gravity remains a major outstanding problem in modern physics.  This fundamental, yet elusive, theory would provide the microscopic structure underlying  systems whose long distance behavior is governed by Einstein's theory of general relativity.  To date, remarkable progress has emerged from concrete realizations~\cite{Maldacena:1997re,Witten:1998qj,Aharony:1999ti,Ryu:2006bv} of the holographic principle \cite{tHooft:1993dmi,Susskind:1994vu,Bousso:2002ju}, underscoring the power of identifying a lower-dimensional non-gravitational description of the same physics. Celestial holography provides a new approach to quantum gravity in asymptotically flat spacetimes by seeking to establish such a holographic correspondence.

More specifically, celestial holography proposes a duality between gravitational scattering in asymptotically flat spacetimes and a conformal field theory living on the  celestial sphere 
 \cite{Strominger:2017zoo,Pasterski:2019ceq,Raclariu:2021zjz,Pasterski:2021rjz}. 
One of the core features of this program is the reorganization of observables according to the principle of putting symmetries front and center. This framework makes manifest infinite dimensional enhancements that are typically hidden in soft theorems~\cite{Strominger:2017zoo}. The central objects of study are celestial amplitudes, which recast the gravitational $\mathcal{S}$-matrix into a basis of boost eigenstates. These transform like the correlation functions of a conformal field theory, providing a key entry in the holographic dictionary. The quest for a self-consistent intrinsically-defined dual theory is thus isomorphic to the $\mathcal{S}$-matrix program~\cite{Eden:1966dnq,Elvang:2013cua,Henn:2014yza}.

In this white paper we review the asymptotic symmetry origins of this program and the basic building blocks of the celestial  holographic dictionary, before presenting some of the prospects for merging tools from the relativity, amplitudes, and bootstrap communities in the form of potential applications and open questions. For concreteness we restrict to massless scattering in 3+1 dimensions throughout, though our story naturally generalizes to both higher dimensions~\cite{Pasterski:2017kqt,Kapec:2017gsg,Banerjee:2019aoy,Banerjee:2019tam} and massive external states~\cite{Pasterski:2017kqt,Pasterski:2016qvg,Law:2020tsg,Muck:2020wtx,Narayanan:2020amh}.

 \section{Symmetries from the Infrared}
 
Matching the symmetries on both sides of the proposed duality is the natural first step towards constructing a holographic dictionary from the ground up. In the bulk the relevant question is: {\it what are the symmetries of asymptotically flat spacetimes?} This question was first tackled by Bondi, van der Burg, Metzner, and Sachs~\cite{Bondi:1962px,Sachs:1962wk,Sachs:1962zza}. Asymptotically flat spacetimes are a class of solutions to Einstein's equations with vanishing cosmological constant and localized matter stress-energy~\cite{Bondi:1962px,Sachs:1962wk,Sachs:1962zza,Ashtekar:1981sf,ctx15533540510006421}, which capture the physics of isolated gravitational systems.  Asymptotic symmetries are diffeomorphisms that preserve the causal structure of the boundary in a conformal compactification.  They can be identified as the residual diffeomorphisms that remain after fixing appropriate gauge and boundary conditions. Namely 
\be\label{asg}
\mathrm{Asymptotic~Symmetries}=\frac{\mathrm{Allowed~Symmetries}}{\mathrm{Trivial~Symmetries}},
\ee
where we quotient by symmetries that act trivially on the phase space. The procedure for identifying the asymptotic symmetries involves a delicate interplay between the allowed boundary conditions and resulting asymptotic symmetry group. For example, BMS employed boundary conditions that led to an enhancement of the Poincar\'e group that includes angle-dependent translations of the generators of null infinity, known as supertranslations.  However, it was not until rather recently that studies considered boundary conditions that were sufficiently relaxed to permit local enhancements of the Lorentz subgroup, called superrotations \cite{Barnich:2011ct,Barnich:2011mi}.

This sensitivity to the choice of boundary conditions lends a degree of arbitrariness that one would like to mitigate. A directly observable consequence of asymptotic symmetries, known as the memory effect, informs this choice, singling out physically relevant solutions and symmetries. The celestial holography program is rooted in the interesting observation~\cite{Strominger:2013lka,Strominger:2013jfa} that the perturbative gravitational $\mathcal{S}$-matrix provides additional guidance in addressing this question.  In particular, the Ward identities for asymptotic symmetries manifest as soft theorems of the $\mathcal{S}$-matrix (see \cite{Strominger:2017zoo} and references therein)
\be
\langle out | Q^+\mathcal{S}-\mathcal{S}Q^-|in\rangle=0~~\Leftrightarrow~~\langle out| a(\omega)\mathcal{S}|in\rangle \underset{\omega \to 0}{ \propto} \langle out|\mathcal{S}|in\rangle.
\ee
  In fact there is an equivalence~\cite{Strominger:2014pwa,Pasterski:2015tva,Pasterski:2015zua} between memory effects~\cite{1974SvA....18...17Z,Braginsky:1986ia,gravmem3,Bieri:2013hqa,Susskind:2015hpa,Nichols:2017rqr,Pate:2017vwa,Ball:2018prg,Himwich:2019qmj}, soft theorems~\cite{Low:1954kd,Low:1958sn,Weinberg:1965nx,Cachazo:2014fwa,Sen:2017nim,Laddha:2017ygw,Laddha:2018myi}, and Ward identities~\cite{He:2014laa,Kapec:2014opa,He:2014cra,Lysov:2014csa,Campiglia:2014yka,Kapec:2014zla,Avery:2015gxa,Strominger:2015bla,Campiglia:2015lxa,Avery:2015rga,Dumitrescu:2015fej,Avery:2015iix,Lysov:2015jrs,Campiglia:2016hvg,Campiglia:2016efb,He:2017fsb,Campiglia:2017dpg,Mitra2017,Laddha:2017vfh,He2018,Kapec2018,Liu:2021dyq} for asymptotic symmetries~\cite{Barnich:2016lyg,Compere:2016jwb,Strominger:2016wns,Compere:2018ylh,Alessio:2019cch,Campiglia:2020qvc,Barnich:2021dta,Freidel:2021cjp,Donnay:2021wrk,Aneesh:2021uzk}.\footnote{For their generalizations to higher dimensions and massive external states, see~\cite{Kapec:2015vwa,Pate:2017fgt,He:2019jjk,He:2019pll,Freidel:2019ohg,Campoleoni:2020ejn} and~\cite{Campiglia:2015qka,Kapec:2015ena,Campiglia:2015kxa}.}
  Moreover, the symmetry generators are naturally represented by currents in a two-dimensional theory \cite{Strominger:2013lka,He:2015zea,Kapec:2016jld,Kapec:2017gsg,Nande:2017dba,Himwich:2019dug,Himwich:2020rro}.  In a particularly rewarding example of applying this equivalence, the aforementioned speculation about the physicality of superrotations led to the discovery of a subleading soft graviton theorem~\cite{Cachazo:2014fwa}.  This in turn established the Ward identity for a Virasoro symmetry~\cite{Kapec:2014opa} and revealed a new memory effect~\cite{Pasterski:2015tva}, as well as a 4D scattering mode that behaves like the stress tensor of a 2D conformal field theory \cite{Kapec:2016jld}. 

 Thus, a foundational lesson that emerges from symmetry is that a holographic dual to quantum gravity in four-dimensional asymptotically flat spacetimes must admit a two-dimensional conformal symmetry! Accordingly, the proposed dual theory built upon this 2D conformal symmetry is referred to as the \emph{celestial conformal field theory} (CCFT).

  \section{Building a Holographic Dictionary}

The subleading soft graviton mode couples to
an external particle of helicity $\ell$ and energy $\omega$ through the operators~\cite{Cachazo:2014fwa,Lysov:2014csa}
\be
\hat{h}=\frac{1}{2}(\ell-\omega\p_\omega),~~~\hat{\bar{h}}=\frac{1}{2}(-\ell-\omega\p_\omega).
\ee
 Scattering states that diagonalize $\hat{h}$ and $\hat{\bar{h}}$ are constructed by transforming from momentum to boost eigenstates~\cite{deBoer:2003vf,Lysov:2014csa,Kapec:2016jld,Cheung:2016iub,Pasterski:2016qvg,Pasterski:2017kqt,Pasterski:2017ylz}. For massless particles, this can be achieved via a Mellin transform in the energy
\be\label{mellin}
\langle \mathcal{O}^\pm_{\Delta_1}(z_1,\bz_1)...\mathcal{O}^\pm_{\Delta_n}(z_n,\bz_n)\rangle=\prod_{i=1}^n \int_0^\infty d\omega_i \omega_i^{\Delta_i-1} \langle out|\mathcal{S}|in\rangle,
\ee
where $(z_i, \bz_i)$ parametrize the direction of propagation of the $i$th particle, $\Delta_i = h_i + \bar{h}_i$ is its scaling dimension or Rindler energy, and the $\pm$ superscripts distinguish between incoming and outgoing operators.  Generalizations of this basis have been explored in~\cite{Pasterski:2017kqt,ss,Fan:2021isc,Atanasov:2021cje,Crawley:2021ivb,Sharma:2021gcz,Kapec:2021eug}.   We thus have a map between traditional $\mathcal{S}$-matrix elements and correlation functions on the celestial sphere. 

The equivalence between Lorentz transformations in four dimensions and global conformal transformations on the celestial sphere guarantees that the operators constructed from Mellin-transformed wavepackets transform as quasi-primaries with conformal weights $(h, \bar{h})$.  In a gravitational theory, these are enhanced to Virasoro primaries via their coupling to the subleading soft graviton \cite{Kapec:2014opa,Kapec:2016jld,Cheung:2016iub,Fotopoulos:2019tpe}.

Given the central correspondence \eqref{mellin}, a more complete holographic dictionary is readily built by directly transforming known amplitudes and features thereof. Universal properties including soft and collinear limits are of particular interest. We present a summary of the current status, organized in terms of standard CFT data, namely the spectrum and OPE coefficients.
  
  \paragraph{The Spectrum} 
  
Finite-energy scattering states, characterized by normalizable single-particle wavepackets in the bulk \cite{Pasterski:2017kqt}, are captured by celestial operators with a spectrum~\cite{deBoer:2003vf} \be\Delta=1+i\lambda,~~\lambda\in\mathbb{R}.\ee Nevertheless, it has proven judicious to analytically continue $\Delta$ over the complex plane~\cite{Donnay:2020guq} in order to incorporate both the action of translations which shift $\Delta \to \Delta+1$~\cite{Donnay:2018neh,Stieberger:2018onx,Law:2019glh}, as well as soft theorems which involve operators with (half)-integer dimensions~\cite{Donnay:2018neh,Fotopoulos:2019tpe,Adamo:2019ipt,Ball:2019atb, Guevara:2019ypd,Pate:2019lpp,Fotopoulos:2020bqj,Guevara:2021abz}.  The conformal dimensions in the latter admit shortened representations of the global conformal symmetry \cite{Banerjee:2018gce,Banerjee:2018fgd,Banerjee:2019aoy,Banerjee:2019tam,Guevara:2021abz,Pasterski:2021fjn,Pasterski:2021dqe}, explaining the observed organization of soft theorems into these multiplets.

   Remarkably, factorization in the soft limit of scattering persists in the boost eigenstate basis~\cite{Cheung:2016iub,Fan:2019emx,Fotopoulos:2019tpe,Pate:2019mfs,Adamo:2019ipt,Puhm:2019zbl,Guevara:2019ypd}. In particular, soft theorems, which prescribe the Laurent coefficients of an expansion about $\omega \to 0$, become statements specifying the residues of simple poles in the corresponding external dimension $\Delta$. More generally, IR and UV phenomena can each have a rather stark and distinct effect on the analyticity of celestial amplitudes in their conformal dimensions~\cite{Stieberger:2018edy,Arkani-Hamed:2020gyp,Chang:2021wvv}.  For example, as a function of the net conformal dimension (which is Mellin-conjugate to the center of mass energy), celestial amplitudes admit a series of simple poles.  The locations of the poles distinguish UV completions of quantum field theories from expected UV completions of quantum gravity and the residues reproduce the coefficients in an EFT expansion.

 \paragraph{OPE Coefficients} 
 The OPE data of the celestial CFT encode the behavior of scattering in collinear limits \cite{Fan:2019emx,Fotopoulos:2019tpe,Pate:2019lpp,Fotopoulos:2019vac,Banerjee:2020kaa,Fan:2020xjj,Ebert:2020nqf,Himwich:2021dau,Guevara:2021tvr}. This follows from the equivalence between the limit in which two operators approach one another on the celestial sphere and the limit in which two massless on-shell momenta become collinear. Celestial OPE coefficients obey symmetry constraints intrinsic to the CCFT. While superrotations establish a Virasoro symmetry naturally incorporated into the standard CFT technology, global translations, as well as the infinity of soft symmetries associated with (half)-integer dimension insertions, imply non-standard constraints on the CFT data. Curiously the Ward identities arising from soft theorems lead to recursion relations for the OPE data~\cite{Pate:2019lpp,Himwich:2021dau} and, given certain analyticity assumptions in $\Delta$, are sufficient to completely bootstrap the leading and, in some cases, subleading collinear behavior of scattering \cite{Pate:2019lpp,Banerjee:2020kaa,Banerjee:2020zlg,Banerjee:2020vnt,Ebert:2020nqf,Banerjee:2021dlm,Himwich:2021dau}. This quite unique feature of CCFT highlights the power of our symmetry-based approach, which apparently allows 4D kinematics to determine the dynamics of the 2D dual. Conversely, there is even recent evidence that celestial symmetries can be used to determine certain bulk dynamics~\cite{Freidel:2021qpz}. \vspace{1em}


 There are many more features of the celestial holographic dictionary that are under active investigation for which the following two approaches have proven fruitful:
 \begin{enumerate}
     \item Determine properties of CCFT from known behavior of scattering amplitudes. 
     \item Establish intrinsic CCFT structure by applying known CFT properties or techniques.
 \end{enumerate}
 In the first category, we have a wide variety of efforts to understand loop effects~\cite{Banerjee:2017jeg,Gonzalez:2020tpi,Albayrak:2020saa,Arkani-Hamed:2020gyp}, double copy relations~\cite{Casali:2020vuy,Casali:2020uvr,Kalyanapuram:2020epb,Pasterski:2020pdk}, supersymmetry~\cite{Muck:2020wtx,Narayanan:2020amh,Pasterski:2020pdk,Fotopoulos:2020bqj,Brandhuber:2021nez,Jiang:2021xzy,Hu:2021lrx,Himwich:2021dau,Ferro:2021dub,Jiang:2021ovh,Pano:2021ewd}, changes in signature~\cite{Atanasov:2021cje,Atanasov:2021oyu,Bhattacharjee:2021mdc}, twistor methods~\cite{Adamo:2019ipt,Sharma:2021gcz,Adamo:2021lrv,Adamo:2021zpw}, and more general amplitudes techniques~\cite{Schreiber:2017jsr,Zlotnikov2018,Ferro:2021dub,Hu:2021lrx,Melton:2021kkz} and features~\cite{Lam:2017ofc,Arkani-Hamed:2020gyp,Chang:2021wvv}.  For the second, there have been extensive investigations of constraints from celestial symmetries~\cite{Stieberger:2018onx,Banerjee:2018gce,Banerjee:2018fgd,Banerjee:2019aoy,Banerjee:2019tam,Law:2019glh,Pate:2019lpp,Law:2020xcf,Banerjee:2020kaa,Banerjee:2020zlg,Pasterski:2021dqe,Pasterski:2021fjn,Guevara:2021abz,Banerjee:2021cly,Strominger:2021lvk,Gupta:2021cwo,Donnay:2021wrk,Banerjee:2021dlm,Banerjee:2020vnt,Jiang:2021csc,Jiang:2021ovh,Banerjee:2021uxe,Adamo:2021lrv,Himwich:2021dau,Ahn:2021erj}, the appropriate representations~\cite{Banerjee:2019prz,Liu:2021tif,Lippstreu:2021avq,Liu:2021xlm,Law:2020tsg}, 2D EFT models for conformally soft sectors~\cite{Cheung:2016iub,Nande:2017dba,Himwich:2020rro,Nguyen:2020hot,Kalyanapuram:2020aya,Magnea:2021fvy,Gonzalez:2021dxw,Kalyanapuram:2021bvf,Nguyen:2021qkt,Kalyanapuram:2021tnl,Kalyanapuram:2021fic,Kalyanapuram:2021dql,Nastase:2021izh}, the celestial state operator correspondence~\cite{Crawley:2021ivb}, and conformal block decomposition~\cite{Nandan:2019jas,Fan:2021isc,Atanasov:2021cje,Fan:2021pbp,Guevara:2021tvr}.  These highlights are only a snapshot of an active venture.  
 While there is a rich variety of open questions regarding technical aspects of the CCFT dictionary, let us turn now to the big picture goals for this program that we hope to accomplish in the near future.
 
 \vspace{1em}

    \section{Applications and Open Questions}

\paragraph{Measuring Memory Effects} 
 The theoretical framework of asymptotic flatness is expected to describe a wide range of experimentally relevant systems from collider to astrophysical (but sub-cosmological) scales. Memory effects are low-frequency observable signatures~\cite{1974SvA....18...17Z,Braginsky:1986ia,gravmem3,Bieri:2013hqa} of symmetry enhancements~\cite{Strominger:2014pwa,Pasterski:2015zua,Pasterski:2015tva,Pate:2017vwa} and thus provide an experimental litmus test for the relevance of the assumed boundary conditions to a given physical process.  The soft physics program has identified new memory effects in both gravity (spin memory~\cite{Pasterski:2015tva}) and gauge theory (color memory~\cite{Pate:2017vwa}), adding to the known examples of leading gravitational~\cite{1974SvA....18...17Z,Braginsky:1986ia,gravmem3,Strominger:2014pwa} and electromagnetic~\cite{Bieri:2013hqa,Pasterski:2015zua,Susskind:2015hpa} memory. In fact, experimental investigations into the leading gravitational memory effect are already underway~\cite{Hubner:2019sly}  and the choice of BMS frame plays an important role in extracting the correct gravitational waveform~\cite{Mitman:2021xkq}. Moreover, color memory is a natural observable \cite{Ball:2018prg} in the physical regime that will be probed by the recently launched electron-ion collider (EIC) project \cite{Accardi:2012qut,AbdulKhalek:2021gbh}. It is tantalizing to ask: Can spin memory be measured by LISA?  Will color memory be observed at the EIC? At the moment there are various proof-of-concept measurement proposals~\cite{Pasterski:2015zua,Pasterski:2015tva,Susskind:2015hpa,Pate:2017vwa} and experimental feasibility studies~\cite{Nichols:2017rqr} for these lesser known memories, but we expect there is much more to come.

\paragraph{Constraining Black Hole Evaporation} The infinite-dimensional symmetries that motivated the celestial holography program have interesting applications in their own right.  In the context of infrared divergences, the notion of asymptotic charge  non-conservation provides a beautiful interpretation of the vanishing amplitude for emitting a finite number of soft particles~\cite{Kapec:2017tkm}. Upon including an infinite number of these soft emissions, one obtains a finite result \emph{precisely because} the contributions from the soft quanta render the asymptotic charges conserved.  By identifying symmetry as the underlying origin of infrared divergences, this perspective establishes a general principle for constructing infrared safe amplitudes~\cite{Chung:1965zza,Kulish:1970ut,Gomez:2016hxz,Kapec:2017tkm,Choi:2017bna,Choi:2017ylo,Gomez:2018war,Choi:2019rlz,Arkani-Hamed:2020gyp,Pasterski:2021dqe}.

The implications of asymptotic symmetry conservation laws are even more rich in the presence of horizons, which can carry soft hair~\cite{Hawking:2015qqa,Hawking:2016msc,Hawking:2016sgy,Strominger:2017aeh}.  For example, asymptotic symmetries constrain black hole evaporation and thereby provide insight into questions about information loss and firewalls~\cite{Donnay:2015abr,Mirbabayi:2016axw,Donnay:2016ejv,Bousso:2017dny,Donnay:2018ckb,Haco:2018ske,Haco:2019ggi,Rahman:2019bmk,Pasterski:2020xvn}. Determining the precise phrasing of these statements within the CCFT framework is an active area of research~\cite{Kapec:2016aqd,Cheung:2016iub,Laddha:2020kvp} which will help bridge the gap between the celestial holography program and standard techniques from AdS holography for quantum information exchange. To this end, it would prove fruitful to understand how to recover celestial CFT from a flat limit of AdS by, for example, zooming into a bulk point~\cite{Hijano:2019qmi,Hijano:2020szl} or relaxing the standard boundary conditions to allow for radiation~\cite{Compere:2019bua}.

\paragraph{A Boo\underline{s}tstrap Program}

The celestial holography program is inherently isomorphic to the  $\mathcal{S}$-matrix program.  Namely any intrinsic definition of the celestial CFT should include a set of rules for constructing consistent (gravitational) scattering matrices as a function of their on-shell data.  Our discussion of the celestial holographic dictionary demonstrates the success with which we have exploited soft and collinear behavior of scattering amplitudes to constrain and in some cases determine properties of celestial amplitudes. While CCFT may be an exotic 2D CFT~\cite{Pasterski:2017ylz,Schreiber:2017jsr}, we have already seen indications of a simple analytic structure in the boost weights that encodes information about bulk physics~\cite{Lam:2017ofc,Arkani-Hamed:2020gyp,Chang:2021wvv}. Recently, the CCFT framework facilitated the identification of a ${\rm w}_{1+\infty}$ algebra underlying the soft symmetries of the $\mathcal{S}$-matrix~\cite{Guevara:2021abz,Strominger:2021lvk,Adamo:2021lrv,Jiang:2021csc,Jiang:2021ovh,Himwich:2021dau}.  In turn, this symmetry algebra supplies the CCFT with yet more powerful organizing principles. Codifying the rules for a consistent CCFT will be an important step for bootstrapping the $\mathcal{S}$-matrix in the boost basis.

 \paragraph{Connecting to String Theory}
A crowning achievement for the celestial holography program would be for it to determine concretely whether string theories are the only consistent theories of (asymptotically flat) quantum gravity. Thus far, it has been observed that even at tree-level, string theoretic amplitudes are sufficiently well-behaved in the UV to admit a celestial representation~\cite{Stieberger:2018edy,Arkani-Hamed:2020gyp}. Yet another interesting line of inquiry is into the relation (and possible equivalence?) between CCFTs and worldsheet constructions of quantum gravity. Recent results matching celestial OPEs to worldsheet OPEs in both standard and twistor string contexts~\cite{Jiang:2021csc,Adamo:2021zpw}, motivated by the discovery of symmetry algebras common to single helicity sectors of CCFT~\cite{Guevara:2021abz,Strominger:2021lvk,Adamo:2021lrv,Jiang:2021csc,Jiang:2021ovh,Himwich:2021dau} and stringy constructions, evoke even more anticipation!

\section*{Acknowledgements}

This subfield has emerged as a collective effort of the many authors presented in the references below.  We are indebted to all of our collaborators and colleagues, especially our phenomenal advisor Andrew Strominger. We thank Laurent Freidel, Andrew Strominger, and Herman Verlinde for comments on a draft. The work of S.P. is supported by the Sam B. Treiman Fellowship at the Princeton Center for Theoretical Science. The work of M.P is supported by the Harvard Society of Fellows, DOE grant de-sc/000787, and the Black Hole Initiative at Harvard University, funded by grants from the John Templeton Foundation and the Gordon and Betty Moore Foundation. The work of A.R. at the Perimeter Institute is supported by the Stephen Hawking fellowship and the Government of Canada through the Department of Innovation, Science and Industry Canada and by the Province of Ontario through the Ministry of Colleges and Universities.

\bibliographystyle{utphys}
\bibliography{references}

\providecommand{\href}[2]{#2}\begingroup\raggedright\begin{thebibliography}{100}

\bibitem{Maldacena:1997re}
J.~M. Maldacena, ``{The Large N limit of superconformal field theories and
  supergravity},'' \href{http://dx.doi.org/10.1023/A:1026654312961}{{\em Adv.
  Theor. Math. Phys.} {\bfseries 2} (1998) 231--252},
  \href{http://arxiv.org/abs/hep-th/9711200}{{\ttfamily arXiv:hep-th/9711200}}.

\bibitem{Witten:1998qj}
E.~Witten, ``{Anti-de Sitter space and holography},''
  \href{http://dx.doi.org/10.4310/ATMP.1998.v2.n2.a2}{{\em Adv. Theor. Math.
  Phys.} {\bfseries 2} (1998) 253--291},
\href{http://arxiv.org/abs/hep-th/9802150}{{\ttfamily arXiv:hep-th/9802150
  [hep-th]}}.

\bibitem{Aharony:1999ti}
O.~Aharony, S.~S. Gubser, J.~M. Maldacena, H.~Ooguri, and Y.~Oz, ``{Large N
  field theories, string theory and gravity},''
  \href{http://dx.doi.org/10.1016/S0370-1573(99)00083-6}{{\em Phys. Rept.}
  {\bfseries 323} (2000) 183--386},
  \href{http://arxiv.org/abs/hep-th/9905111}{{\ttfamily arXiv:hep-th/9905111}}.

\bibitem{Ryu:2006bv}
S.~Ryu and T.~Takayanagi, ``{Holographic derivation of entanglement entropy
  from AdS/CFT},'' \href{http://dx.doi.org/10.1103/PhysRevLett.96.181602}{{\em
  Phys. Rev. Lett.} {\bfseries 96} (2006) 181602},
  \href{http://arxiv.org/abs/hep-th/0603001}{{\ttfamily arXiv:hep-th/0603001}}.

\bibitem{tHooft:1993dmi}
G.~'t~Hooft, ``{Dimensional reduction in quantum gravity},'' {\em Conf. Proc.
  C} {\bfseries 930308} (1993) 284--296,
  \href{http://arxiv.org/abs/gr-qc/9310026}{{\ttfamily arXiv:gr-qc/9310026}}.

\bibitem{Susskind:1994vu}
L.~Susskind, ``{The World as a hologram},''
  \href{http://dx.doi.org/10.1063/1.531249}{{\em J. Math. Phys.} {\bfseries 36}
  (1995) 6377--6396}, \href{http://arxiv.org/abs/hep-th/9409089}{{\ttfamily
  arXiv:hep-th/9409089}}.

\bibitem{Bousso:2002ju}
R.~Bousso, ``{The Holographic principle},''
  \href{http://dx.doi.org/10.1103/RevModPhys.74.825}{{\em Rev. Mod. Phys.}
  {\bfseries 74} (2002) 825--874},
  \href{http://arxiv.org/abs/hep-th/0203101}{{\ttfamily arXiv:hep-th/0203101}}.

\bibitem{Strominger:2017zoo}
A.~Strominger, {\em {Lectures on the Infrared Structure of Gravity and Gauge
  Theory}}.
\newblock {Princeton University Press}, 2018.
\newblock
\href{http://arxiv.org/abs/1703.05448}{{\ttfamily arXiv:1703.05448 [hep-th]}}.
\newblock

\bibitem{Pasterski:2019ceq}
S.~Pasterski, ``{Implications of Superrotations},''
  \href{http://dx.doi.org/10.1016/j.physrep.2019.09.006}{{\em Phys. Rept.}
  {\bfseries 829} (2019) 1--35},
  \href{http://arxiv.org/abs/1905.10052}{{\ttfamily arXiv:1905.10052
  [hep-th]}}.

\bibitem{Raclariu:2021zjz}
A.-M. Raclariu, ``{Lectures on Celestial Holography},''
  \href{http://arxiv.org/abs/2107.02075}{{\ttfamily arXiv:2107.02075
  [hep-th]}}.

\bibitem{Pasterski:2021rjz}
S.~Pasterski, ``{Lectures on Celestial Amplitudes},''
  \href{http://arxiv.org/abs/2108.04801}{{\ttfamily arXiv:2108.04801
  [hep-th]}}.

\bibitem{Eden:1966dnq}
R.~J. Eden, P.~V. Landshoff, D.~I. Olive, and J.~C. Polkinghorne, {\em {The
  analytic S-matrix}}.
\newblock Cambridge Univ. Press, Cambridge, 1966.

\bibitem{Elvang:2013cua}
H.~Elvang and Y.-t. Huang, ``{Scattering Amplitudes},''
  \href{http://arxiv.org/abs/1308.1697}{{\ttfamily arXiv:1308.1697 [hep-th]}}.

\bibitem{Henn:2014yza}
J.~M. Henn and J.~C. Plefka,
  \href{http://dx.doi.org/10.1007/978-3-642-54022-6}{{\em {Scattering
  Amplitudes in Gauge Theories}}}, vol.~883.
\newblock Springer, Berlin, 2014.

\bibitem{Pasterski:2017kqt}
S.~Pasterski and S.-H. Shao, ``{Conformal basis for flat space amplitudes},''
  \href{http://dx.doi.org/10.1103/PhysRevD.96.065022}{{\em Phys. Rev.}
  {\bfseries D96} no.~6, (2017) 065022},
\href{http://arxiv.org/abs/1705.01027}{{\ttfamily arXiv:1705.01027 [hep-th]}}.

\bibitem{Kapec:2017gsg}
D.~Kapec and P.~Mitra, ``{A $d$-Dimensional Stress Tensor for Mink$_{d+2}$
  Gravity},'' \href{http://dx.doi.org/10.1007/JHEP05(2018)186}{{\em JHEP}
  {\bfseries 05} (2018) 186},
\href{http://arxiv.org/abs/1711.04371}{{\ttfamily arXiv:1711.04371 [hep-th]}}.

\bibitem{Banerjee:2019aoy}
S.~Banerjee, P.~Pandey, and P.~Paul, ``{Conformal properties of soft operators:
  Use of null states},''
  \href{http://dx.doi.org/10.1103/PhysRevD.101.106014}{{\em Phys. Rev. D}
  {\bfseries 101} no.~10, (2020) 106014},
  \href{http://arxiv.org/abs/1902.02309}{{\ttfamily arXiv:1902.02309
  [hep-th]}}.

\bibitem{Banerjee:2019tam}
S.~Banerjee and P.~Pandey, ``{Conformal properties of soft-operators. Part II.
  Use of null-states},'' \href{http://dx.doi.org/10.1007/JHEP02(2020)067}{{\em
  JHEP} {\bfseries 02} (2020) 067},
  \href{http://arxiv.org/abs/1906.01650}{{\ttfamily arXiv:1906.01650
  [hep-th]}}.

\bibitem{Pasterski:2016qvg}
S.~Pasterski, S.-H. Shao, and A.~Strominger, ``{Flat Space Amplitudes and
  Conformal Symmetry of the Celestial Sphere},''
  \href{http://dx.doi.org/10.1103/PhysRevD.96.065026}{{\em Phys. Rev.}
  {\bfseries D96} no.~6, (2017) 065026},
\href{http://arxiv.org/abs/1701.00049}{{\ttfamily arXiv:1701.00049 [hep-th]}}.

\bibitem{Law:2020tsg}
Y.~A. Law and M.~Zlotnikov, ``{Massive Spinning Bosons on the Celestial
  Sphere},'' \href{http://dx.doi.org/10.1007/JHEP06(2020)079}{{\em JHEP}
  {\bfseries 06} (2020) 079}, \href{http://arxiv.org/abs/2004.04309}{{\ttfamily
  arXiv:2004.04309 [hep-th]}}.

\bibitem{Muck:2020wtx}
L.~Iacobacci and W.~M\"uck, ``{Conformal Primary Basis for Dirac Spinors},''
  \href{http://dx.doi.org/10.1103/PhysRevD.102.106025}{{\em Phys. Rev. D}
  {\bfseries 102} no.~10, (2020) 106025},
  \href{http://arxiv.org/abs/2009.02938}{{\ttfamily arXiv:2009.02938
  [hep-th]}}.

\bibitem{Narayanan:2020amh}
S.~A. Narayanan, ``{Massive Celestial Fermions},''
  \href{http://arxiv.org/abs/2009.03883}{{\ttfamily arXiv:2009.03883
  [hep-th]}}.

\bibitem{Bondi:1962px}
H.~Bondi, M.~G.~J. van~der Burg, and A.~W.~K. Metzner, ``{Gravitational waves
  in general relativity. 7. Waves from axisymmetric isolated systems},''
\href{http://dx.doi.org/10.1098/rspa.1962.0161}{{\em Proc. Roy. Soc. Lond.}
  {\bfseries A269} (1962) 21--52}.

\bibitem{Sachs:1962wk}
R.~K. Sachs, ``{Gravitational waves in general relativity. 8. Waves in
  asymptotically flat space-times},''
\href{http://dx.doi.org/10.1098/rspa.1962.0206}{{\em Proc. Roy. Soc. Lond.}
  {\bfseries A270} (1962) 103--126}.

\bibitem{Sachs:1962zza}
R.~Sachs, ``{Asymptotic symmetries in gravitational theory},''
\href{http://dx.doi.org/10.1103/PhysRev.128.2851}{{\em Phys. Rev.} {\bfseries
  128} (1962) 2851--2864}.

\bibitem{Ashtekar:1981sf}
A.~Ashtekar, ``{Asymptotic Quantization of the Gravitational Field},''
\href{http://dx.doi.org/10.1103/PhysRevLett.46.573}{{\em Phys. Rev. Lett.}
  {\bfseries 46} (1981) 573--576}.

\bibitem{ctx15533540510006421}
D.~Christodoulou, {\em The Global Nonlinear Stability of the Minkowski Space
  (PMS-41)}.
\newblock Princeton University Press, Princeton, 2014-07-14.

\bibitem{Barnich:2011ct}
G.~Barnich and C.~Troessaert, ``{Supertranslations call for superrotations},''
  {\em PoS} {\bfseries CNCFG} (2010) 010,
  \href{http://arxiv.org/abs/1102.4632}{{\ttfamily arXiv:1102.4632 [gr-qc]}}.
[Ann. U. Craiova Phys.21, S11 (2011)].

\bibitem{Barnich:2011mi}
G.~Barnich and C.~Troessaert, ``{BMS charge algebra},''
  \href{http://dx.doi.org/10.1007/JHEP12(2011)105}{{\em JHEP} {\bfseries 12}
  (2011) 105},
\href{http://arxiv.org/abs/1106.0213}{{\ttfamily arXiv:1106.0213 [hep-th]}}.

\bibitem{Strominger:2013lka}
A.~Strominger, ``{Asymptotic Symmetries of Yang-Mills Theory},''
  \href{http://dx.doi.org/10.1007/JHEP07(2014)151}{{\em JHEP} {\bfseries 07}
  (2014) 151},
\href{http://arxiv.org/abs/1308.0589}{{\ttfamily arXiv:1308.0589 [hep-th]}}.

\bibitem{Strominger:2013jfa}
A.~Strominger, ``{On BMS Invariance of Gravitational Scattering},''
  \href{http://dx.doi.org/10.1007/JHEP07(2014)152}{{\em JHEP} {\bfseries 07}
  (2014) 152},
\href{http://arxiv.org/abs/1312.2229}{{\ttfamily arXiv:1312.2229 [hep-th]}}.

\bibitem{Strominger:2014pwa}
A.~Strominger and A.~Zhiboedov, ``{Gravitational Memory, BMS Supertranslations
  and Soft Theorems},'' \href{http://dx.doi.org/10.1007/JHEP01(2016)086}{{\em
  JHEP} {\bfseries 01} (2016) 086},
  \href{http://arxiv.org/abs/1411.5745}{{\ttfamily arXiv:1411.5745 [hep-th]}}.

\bibitem{Pasterski:2015tva}
S.~Pasterski, A.~Strominger, and A.~Zhiboedov, ``{New Gravitational
  Memories},'' \href{http://dx.doi.org/10.1007/JHEP12(2016)053}{{\em JHEP}
  {\bfseries 12} (2016) 053}, \href{http://arxiv.org/abs/1502.06120}{{\ttfamily
  arXiv:1502.06120 [hep-th]}}.

\bibitem{Pasterski:2015zua}
S.~Pasterski, ``{Asymptotic Symmetries and Electromagnetic Memory},''
  \href{http://dx.doi.org/10.1007/JHEP09(2017)154}{{\em JHEP} {\bfseries 09}
  (2017) 154},
\href{http://arxiv.org/abs/1505.00716}{{\ttfamily arXiv:1505.00716 [hep-th]}}.

\bibitem{1974SvA....18...17Z}
Y.~B. {Zel'dovich} and A.~G. {Polnarev}, ``{Radiation of gravitational waves by
  a cluster of superdense stars},'' {\em Sov. Astron. Lett} {\bfseries 18}
  (1974) 17.

\bibitem{Braginsky:1986ia}
V.~B. Braginsky and L.~P. Grishchuk, ``{Kinematic Resonance and Memory Effect
  in Free Mass Gravitational Antennas},'' {\em Sov. Phys. JETP} {\bfseries 62}
  (1985) 427--430.
[Zh. Eksp. Teor. Fiz.89,744(1985)].

\bibitem{gravmem3}
V.~B. Braginsky and K.~S. Thorne, ``Gravitational-wave bursts with memory and
  experimental prospects,'' \href{http://dx.doi.org/10.1038/327123a0}{{\em
  Nature} {\bfseries 327} no.~6118, (1987) 123--125}.

\bibitem{Bieri:2013hqa}
L.~Bieri and D.~Garfinkle, ``{An electromagnetic analogue of gravitational wave
  memory},'' \href{http://dx.doi.org/10.1088/0264-9381/30/19/195009}{{\em
  Class. Quant. Grav.} {\bfseries 30} (2013) 195009},
\href{http://arxiv.org/abs/1307.5098}{{\ttfamily arXiv:1307.5098 [gr-qc]}}.

\bibitem{Susskind:2015hpa}
L.~Susskind, ``{Electromagnetic Memory},''
\href{http://arxiv.org/abs/1507.02584}{{\ttfamily arXiv:1507.02584 [hep-th]}}.

\bibitem{Nichols:2017rqr}
D.~A. Nichols, ``{Spin memory effect for compact binaries in the post-Newtonian
  approximation},'' \href{http://dx.doi.org/10.1103/PhysRevD.95.084048}{{\em
  Phys. Rev. D} {\bfseries 95} no.~8, (2017) 084048},
  \href{http://arxiv.org/abs/1702.03300}{{\ttfamily arXiv:1702.03300 [gr-qc]}}.

\bibitem{Pate:2017vwa}
M.~Pate, A.-M. Raclariu, and A.~Strominger, ``{Color Memory: A Yang-Mills
  Analog of Gravitational Wave Memory},''
  \href{http://dx.doi.org/10.1103/PhysRevLett.119.261602}{{\em Phys. Rev.
  Lett.} {\bfseries 119} no.~26, (2017) 261602},
  \href{http://arxiv.org/abs/1707.08016}{{\ttfamily arXiv:1707.08016
  [hep-th]}}.

\bibitem{Ball:2018prg}
A.~Ball, M.~Pate, A.-M. Raclariu, A.~Strominger, and R.~Venugopalan,
  ``{Measuring color memory in a color glass condensate at
  electron\textendash{}ion colliders},''
  \href{http://dx.doi.org/10.1016/j.aop.2019.04.010}{{\em Annals Phys.}
  {\bfseries 407} (2019) 15--28},
  \href{http://arxiv.org/abs/1805.12224}{{\ttfamily arXiv:1805.12224
  [hep-ph]}}.

\bibitem{Himwich:2019qmj}
E.~Himwich, Z.~Mirzaiyan, and S.~Pasterski, ``{A Note on the Subleading Soft
  Graviton},''
\href{http://arxiv.org/abs/1902.01840}{{\ttfamily arXiv:1902.01840 [hep-th]}}.

\bibitem{Low:1954kd}
F.~E. Low, ``{Scattering of light of very low frequency by systems of spin
  1/2},'' \href{http://dx.doi.org/10.1103/PhysRev.96.1428}{{\em Phys. Rev.}
  {\bfseries 96} (1954) 1428--1432}.

\bibitem{Low:1958sn}
F.~E. Low, ``{Bremsstrahlung of very low-energy quanta in elementary particle
  collisions},'' \href{http://dx.doi.org/10.1103/PhysRev.110.974}{{\em Phys.
  Rev.} {\bfseries 110} (1958) 974--977}.

\bibitem{Weinberg:1965nx}
S.~Weinberg, ``{Infrared photons and gravitons},''
\href{http://dx.doi.org/10.1103/PhysRev.140.B516}{{\em Phys. Rev.} {\bfseries
  140} (1965) B516--B524}.

\bibitem{Cachazo:2014fwa}
F.~Cachazo and A.~Strominger, ``{Evidence for a New Soft Graviton Theorem},''
\href{http://arxiv.org/abs/1404.4091}{{\ttfamily arXiv:1404.4091 [hep-th]}}.

\bibitem{Sen:2017nim}
A.~Sen, ``{Subleading Soft Graviton Theorem for Loop Amplitudes},''
  \href{http://dx.doi.org/10.1007/JHEP11(2017)123}{{\em JHEP} {\bfseries 11}
  (2017) 123}, \href{http://arxiv.org/abs/1703.00024}{{\ttfamily
  arXiv:1703.00024 [hep-th]}}.

\bibitem{Laddha:2017ygw}
A.~Laddha and A.~Sen, ``{Sub-subleading Soft Graviton Theorem in Generic
  Theories of Quantum Gravity},''
  \href{http://dx.doi.org/10.1007/JHEP10(2017)065}{{\em JHEP} {\bfseries 10}
  (2017) 065}, \href{http://arxiv.org/abs/1706.00759}{{\ttfamily
  arXiv:1706.00759 [hep-th]}}.

\bibitem{Laddha:2018myi}
A.~Laddha and A.~Sen, ``{Logarithmic Terms in the Soft Expansion in Four
  Dimensions},'' \href{http://dx.doi.org/10.1007/JHEP10(2018)056}{{\em JHEP}
  {\bfseries 10} (2018) 056}, \href{http://arxiv.org/abs/1804.09193}{{\ttfamily
  arXiv:1804.09193 [hep-th]}}.

\bibitem{He:2014laa}
T.~He, V.~Lysov, P.~Mitra, and A.~Strominger, ``{BMS supertranslations and
  Weinberg's soft graviton theorem},''
  \href{http://dx.doi.org/10.1007/JHEP05(2015)151}{{\em JHEP} {\bfseries 05}
  (2015) 151},
\href{http://arxiv.org/abs/1401.7026}{{\ttfamily arXiv:1401.7026 [hep-th]}}.

\bibitem{Kapec:2014opa}
D.~Kapec, V.~Lysov, S.~Pasterski, and A.~Strominger, ``{Semiclassical Virasoro
  symmetry of the quantum gravity $ \mathcal{S}$-matrix},''
  \href{http://dx.doi.org/10.1007/JHEP08(2014)058}{{\em JHEP} {\bfseries 08}
  (2014) 058},
\href{http://arxiv.org/abs/1406.3312}{{\ttfamily arXiv:1406.3312 [hep-th]}}.

\bibitem{He:2014cra}
T.~He, P.~Mitra, A.~P. Porfyriadis, and A.~Strominger, ``{New Symmetries of
  Massless QED},'' \href{http://dx.doi.org/10.1007/JHEP10(2014)112}{{\em JHEP}
  {\bfseries 10} (2014) 112},
\href{http://arxiv.org/abs/1407.3789}{{\ttfamily arXiv:1407.3789 [hep-th]}}.

\bibitem{Lysov:2014csa}
V.~Lysov, S.~Pasterski, and A.~Strominger, ``{Low's Subleading Soft Theorem as
  a Symmetry of QED},''
  \href{http://dx.doi.org/10.1103/PhysRevLett.113.111601}{{\em Phys. Rev.
  Lett.} {\bfseries 113} no.~11, (2014) 111601},
\href{http://arxiv.org/abs/1407.3814}{{\ttfamily arXiv:1407.3814 [hep-th]}}.

\bibitem{Campiglia:2014yka}
M.~Campiglia and A.~Laddha, ``{Asymptotic symmetries and subleading soft
  graviton theorem},'' \href{http://dx.doi.org/10.1103/PhysRevD.90.124028}{{\em
  Phys. Rev.} {\bfseries D90} no.~12, (2014) 124028},
\href{http://arxiv.org/abs/1408.2228}{{\ttfamily arXiv:1408.2228 [hep-th]}}.

\bibitem{Kapec:2014zla}
D.~Kapec, V.~Lysov, and A.~Strominger, ``{Asymptotic Symmetries of Massless QED
  in Even Dimensions},''
  \href{http://dx.doi.org/10.4310/ATMP.2017.v21.n7.a6}{{\em Adv. Theor. Math.
  Phys.} {\bfseries 21} (2017) 1747--1767},
\href{http://arxiv.org/abs/1412.2763}{{\ttfamily arXiv:1412.2763 [hep-th]}}.

\bibitem{Avery:2015gxa}
S.~G. Avery and B.~U.~W. Schwab, ``{Burg-Metzner-Sachs symmetry, string theory,
  and soft theorems},''
  \href{http://dx.doi.org/10.1103/PhysRevD.93.026003}{{\em Phys.\ Rev.\ D}
  {\bfseries 93} (2016) 026003},
  \href{http://arxiv.org/abs/1506.05789}{{\ttfamily arXiv:1506.05789
  [hep-th]}}.

\bibitem{Strominger:2015bla}
A.~Strominger, ``{Magnetic Corrections to the Soft Photon Theorem},''
  \href{http://dx.doi.org/10.1103/PhysRevLett.116.031602}{{\em Phys. Rev.
  Lett.} {\bfseries 116} no.~3, (2016) 031602},
\href{http://arxiv.org/abs/1509.00543}{{\ttfamily arXiv:1509.00543 [hep-th]}}.

\bibitem{Campiglia:2015lxa}
M.~Campiglia, ``{Null to time-like infinity Green's functions for asymptotic
  symmetries in Minkowski spacetime},''
  \href{http://dx.doi.org/10.1007/JHEP11(2015)160}{{\em JHEP} {\bfseries 11}
  (2015) 160},
\href{http://arxiv.org/abs/1509.01408}{{\ttfamily arXiv:1509.01408 [hep-th]}}.

\bibitem{Avery:2015rga}
S.~G. Avery and B.~U.~W. Schwab, ``{Noether\textquoteright{}s second theorem
  and Ward identities for gauge symmetries},''
  \href{http://dx.doi.org/10.1007/JHEP02(2016)031}{{\em JHEP} {\bfseries 02}
  (2016) 031}, \href{http://arxiv.org/abs/1510.07038}{{\ttfamily
  arXiv:1510.07038 [hep-th]}}.

\bibitem{Dumitrescu:2015fej}
T.~T. Dumitrescu, T.~He, P.~Mitra, and A.~Strominger, ``{Infinite-Dimensional
  Fermionic Symmetry in Supersymmetric Gauge Theories},''
  \href{http://arxiv.org/abs/1511.07429}{{\ttfamily arXiv:1511.07429
  [hep-th]}}.

\bibitem{Avery:2015iix}
S.~G. Avery and B.~U. Schwab, ``{Residual Local Supersymmetry and the Soft
  Gravitino},'' \href{http://dx.doi.org/10.1103/PhysRevLett.116.171601}{{\em
  Phys. Rev. Lett.} {\bfseries 116} no.~17, (2016) 171601},
  \href{http://arxiv.org/abs/1512.02657}{{\ttfamily arXiv:1512.02657
  [hep-th]}}.

\bibitem{Lysov:2015jrs}
V.~Lysov, ``{Asymptotic Fermionic Symmetry From Soft Gravitino Theorem},''
  \href{http://arxiv.org/abs/1512.03015}{{\ttfamily arXiv:1512.03015
  [hep-th]}}.

\bibitem{Campiglia:2016hvg}
M.~Campiglia and A.~Laddha, ``{Subleading soft photons and large gauge
  transformations},'' \href{http://dx.doi.org/10.1007/JHEP11(2016)012}{{\em
  JHEP} {\bfseries 11} (2016) 012},
\href{http://arxiv.org/abs/1605.09677}{{\ttfamily arXiv:1605.09677 [hep-th]}}.

\bibitem{Campiglia:2016efb}
M.~Campiglia and A.~Laddha, ``{Sub-subleading soft gravitons and large
  diffeomorphisms},'' \href{http://dx.doi.org/10.1007/JHEP01(2017)036}{{\em
  JHEP} {\bfseries 01} (2017) 036},
\href{http://arxiv.org/abs/1608.00685}{{\ttfamily arXiv:1608.00685 [gr-qc]}}.

\bibitem{He:2017fsb}
T.~He, D.~Kapec, A.-M. Raclariu, and A.~Strominger, ``{Loop-Corrected Virasoro
  Symmetry of 4D Quantum Gravity},''
  \href{http://dx.doi.org/10.1007/JHEP08(2017)050}{{\em JHEP} {\bfseries 08}
  (2017) 050},
\href{http://arxiv.org/abs/1701.00496}{{\ttfamily arXiv:1701.00496 [hep-th]}}.

\bibitem{Campiglia:2017dpg}
M.~Campiglia, L.~Coito, and S.~Mizera, ``{Can scalars have asymptotic
  symmetries?},'' \href{http://dx.doi.org/10.1103/PhysRevD.97.046002}{{\em
  Phys. Rev. D} {\bfseries 97} no.~4, (2018) 046002},
  \href{http://arxiv.org/abs/1703.07885}{{\ttfamily arXiv:1703.07885
  [hep-th]}}.

\bibitem{Mitra2017}
P.~Mitra, {\em {Asymptotic Symmetries in Four-Dimensional Gauge and Gravity
  Theories}}.
\newblock PhD thesis, Harvard U. (main), 5, 2017.

\bibitem{Laddha:2017vfh}
A.~Laddha and P.~Mitra, ``{Asymptotic Symmetries and Subleading Soft Photon
  Theorem in Effective Field Theories},''
  \href{http://dx.doi.org/10.1007/JHEP05(2018)132}{{\em JHEP} {\bfseries 05}
  (2018) 132}, \href{http://arxiv.org/abs/1709.03850}{{\ttfamily
  arXiv:1709.03850 [hep-th]}}.

\bibitem{He2018}
T.~M. He, {\em {On Soft Theorems and Asymptotic Symmetries in Four
  Dimensions}}.
\newblock PhD thesis, Harvard U. (main), 2018.

\bibitem{Kapec2018}
D.~S. Kapec, {\em {Aspects of Symmetry in Asymptotically Flat Spacetimes}}.
\newblock PhD thesis, Harvard U., 5, 2018.

\bibitem{Liu:2021dyq}
Z.~Liu and P.~Mao, ``{Infrared photons and asymptotic symmetries},''
  \href{http://dx.doi.org/10.1016/j.physletb.2021.136698}{{\em Phys. Lett. B}
  {\bfseries 822} (2021) 136698},
  \href{http://arxiv.org/abs/2107.03240}{{\ttfamily arXiv:2107.03240
  [hep-th]}}.

\bibitem{Barnich:2016lyg}
G.~Barnich and C.~Troessaert, ``{Finite BMS transformations},''
  \href{http://dx.doi.org/10.1007/JHEP03(2016)167}{{\em JHEP} {\bfseries 03}
  (2016) 167},
\href{http://arxiv.org/abs/1601.04090}{{\ttfamily arXiv:1601.04090 [gr-qc]}}.

\bibitem{Compere:2016jwb}
G.~Comp\`ere and J.~Long, ``{Vacua of the gravitational field},''
  \href{http://dx.doi.org/10.1007/JHEP07(2016)137}{{\em JHEP} {\bfseries 07}
  (2016) 137}, \href{http://arxiv.org/abs/1601.04958}{{\ttfamily
  arXiv:1601.04958 [hep-th]}}.

\bibitem{Strominger:2016wns}
A.~Strominger and A.~Zhiboedov, ``{Superrotations and Black Hole Pair
  Creation},'' \href{http://dx.doi.org/10.1088/1361-6382/aa5b5f}{{\em Class.
  Quant. Grav.} {\bfseries 34} no.~6, (2017) 064002},
  \href{http://arxiv.org/abs/1610.00639}{{\ttfamily arXiv:1610.00639
  [hep-th]}}.

\bibitem{Compere:2018ylh}
G.~Comp\`ere, A.~Fiorucci, and R.~Ruzziconi, ``{Superboost transitions,
  refraction memory and super-Lorentz charge algebra},''
  \href{http://dx.doi.org/10.1007/JHEP11(2018)200}{{\em JHEP} {\bfseries 11}
  (2018) 200}, \href{http://arxiv.org/abs/1810.00377}{{\ttfamily
  arXiv:1810.00377 [hep-th]}}.

\bibitem{Alessio:2019cch}
F.~Alessio and M.~Arzano, ``{A fuzzy bipolar celestial sphere},''
  \href{http://dx.doi.org/10.1007/JHEP07(2019)028}{{\em JHEP} {\bfseries 07}
  (2019) 028}, \href{http://arxiv.org/abs/1901.01167}{{\ttfamily
  arXiv:1901.01167 [gr-qc]}}.

\bibitem{Campiglia:2020qvc}
M.~Campiglia and J.~Peraza, ``{Generalized BMS charge algebra},''
  \href{http://arxiv.org/abs/2002.06691}{{\ttfamily arXiv:2002.06691 [gr-qc]}}.

\bibitem{Barnich:2021dta}
G.~Barnich and R.~Ruzziconi, ``{Coadjoint representation of the BMS group on
  celestial Riemann surfaces},''
  \href{http://arxiv.org/abs/2103.11253}{{\ttfamily arXiv:2103.11253 [gr-qc]}}.

\bibitem{Freidel:2021cjp}
L.~Freidel, R.~Oliveri, D.~Pranzetti, and S.~Speziale, ``{Extended corner
  symmetry, charge bracket and Einstein's equations},''
  \href{http://dx.doi.org/10.1007/JHEP09(2021)083}{{\em JHEP} {\bfseries 09}
  (2021) 083}, \href{http://arxiv.org/abs/2104.12881}{{\ttfamily
  arXiv:2104.12881 [hep-th]}}.

\bibitem{Donnay:2021wrk}
L.~Donnay and R.~Ruzziconi, ``{BMS Flux Algebra in Celestial Holography},''
  \href{http://arxiv.org/abs/2108.11969}{{\ttfamily arXiv:2108.11969
  [hep-th]}}.

\bibitem{Aneesh:2021uzk}
P.~B. Aneesh, G.~Comp\`ere, L.~P. de~Gioia, I.~Mol, and B.~Swidler,
  ``{Celestial Holography: Lectures on Asymptotic Symmetries},''
  \href{http://arxiv.org/abs/2109.00997}{{\ttfamily arXiv:2109.00997
  [hep-th]}}.

\bibitem{Kapec:2015vwa}
D.~Kapec, V.~Lysov, S.~Pasterski, and A.~Strominger, ``{Higher-dimensional
  supertranslations and Weinberg\textquoteright{}s soft graviton theorem},''
  \href{http://dx.doi.org/10.4310/AMSA.2017.v2.n1.a2}{{\em Ann. Math. Sci.
  Appl.} {\bfseries 02} (2017) 69--94},
  \href{http://arxiv.org/abs/1502.07644}{{\ttfamily arXiv:1502.07644 [gr-qc]}}.

\bibitem{Pate:2017fgt}
M.~Pate, A.-M. Raclariu, and A.~Strominger, ``{Gravitational Memory in Higher
  Dimensions},'' \href{http://dx.doi.org/10.1007/JHEP06(2018)138}{{\em JHEP}
  {\bfseries 06} (2018) 138}, \href{http://arxiv.org/abs/1712.01204}{{\ttfamily
  arXiv:1712.01204 [hep-th]}}.

\bibitem{He:2019jjk}
T.~He and P.~Mitra, ``{Asymptotic symmetries and Weinberg\textquoteright{}s
  soft photon theorem in Mink$_{d+2}$},''
  \href{http://dx.doi.org/10.1007/JHEP10(2019)213}{{\em JHEP} {\bfseries 10}
  (2019) 213}, \href{http://arxiv.org/abs/1903.02608}{{\ttfamily
  arXiv:1903.02608 [hep-th]}}.

\bibitem{He:2019pll}
T.~He and P.~Mitra, ``{Asymptotic symmetries in (d + 2)-dimensional gauge
  theories},'' \href{http://dx.doi.org/10.1007/JHEP10(2019)277}{{\em JHEP}
  {\bfseries 10} (2019) 277}, \href{http://arxiv.org/abs/1903.03607}{{\ttfamily
  arXiv:1903.03607 [hep-th]}}.

\bibitem{Freidel:2019ohg}
L.~Freidel, F.~Hopfm\"uller, and A.~Riello, ``{Asymptotic Renormalization in
  Flat Space: Symplectic Potential and Charges of Electromagnetism},''
  \href{http://dx.doi.org/10.1007/JHEP10(2019)126}{{\em JHEP} {\bfseries 10}
  (2019) 126}, \href{http://arxiv.org/abs/1904.04384}{{\ttfamily
  arXiv:1904.04384 [hep-th]}}.

\bibitem{Campoleoni:2020ejn}
A.~Campoleoni, D.~Francia, and C.~Heissenberg, ``{On asymptotic symmetries in
  higher dimensions for any spin},''
  \href{http://dx.doi.org/10.1007/JHEP12(2020)129}{{\em JHEP} {\bfseries 12}
  (2020) 129}, \href{http://arxiv.org/abs/2011.04420}{{\ttfamily
  arXiv:2011.04420 [hep-th]}}.

\bibitem{Campiglia:2015qka}
M.~Campiglia and A.~Laddha, ``{Asymptotic symmetries of QED and Weinberg's soft
  photon theorem},'' \href{http://dx.doi.org/10.1007/JHEP07(2015)115}{{\em
  JHEP} {\bfseries 07} (2015) 115},
\href{http://arxiv.org/abs/1505.05346}{{\ttfamily arXiv:1505.05346 [hep-th]}}.

\bibitem{Kapec:2015ena}
D.~Kapec, M.~Pate, and A.~Strominger, ``{New Symmetries of QED},''
\href{http://arxiv.org/abs/1506.02906}{{\ttfamily arXiv:1506.02906 [hep-th]}}.

\bibitem{Campiglia:2015kxa}
M.~Campiglia and A.~Laddha, ``{Asymptotic symmetries of gravity and soft
  theorems for massive particles},''
  \href{http://dx.doi.org/10.1007/JHEP12(2015)094}{{\em JHEP} {\bfseries 12}
  (2015) 094}, \href{http://arxiv.org/abs/1509.01406}{{\ttfamily
  arXiv:1509.01406 [hep-th]}}.

\bibitem{He:2015zea}
T.~He, P.~Mitra, and A.~Strominger, ``{2D Kac-Moody Symmetry of 4D Yang-Mills
  Theory},'' \href{http://dx.doi.org/10.1007/JHEP10(2016)137}{{\em JHEP}
  {\bfseries 10} (2016) 137},
\href{http://arxiv.org/abs/1503.02663}{{\ttfamily arXiv:1503.02663 [hep-th]}}.

\bibitem{Kapec:2016jld}
D.~Kapec, P.~Mitra, A.-M. Raclariu, and A.~Strominger, ``{2D Stress Tensor for
  4D Gravity},'' \href{http://dx.doi.org/10.1103/PhysRevLett.119.121601}{{\em
  Phys. Rev. Lett.} {\bfseries 119} no.~12, (2017) 121601},
\href{http://arxiv.org/abs/1609.00282}{{\ttfamily arXiv:1609.00282 [hep-th]}}.

\bibitem{Nande:2017dba}
A.~Nande, M.~Pate, and A.~Strominger, ``{Soft Factorization in QED from 2D
  Kac-Moody Symmetry},'' \href{http://dx.doi.org/10.1007/JHEP02(2018)079}{{\em
  JHEP} {\bfseries 02} (2018) 079},
\href{http://arxiv.org/abs/1705.00608}{{\ttfamily arXiv:1705.00608 [hep-th]}}.

\bibitem{Himwich:2019dug}
E.~Himwich and A.~Strominger, ``{Celestial Current Algebra from Low's
  Subleading Soft Theorem},''
\href{http://arxiv.org/abs/1901.01622}{{\ttfamily arXiv:1901.01622 [hep-th]}}.

\bibitem{Himwich:2020rro}
E.~Himwich, S.~A. Narayanan, M.~Pate, N.~Paul, and A.~Strominger, ``{The Soft
  $\mathcal{S}$-Matrix in Gravity},''
  \href{http://dx.doi.org/10.1007/JHEP09(2020)129}{{\em JHEP} {\bfseries 09}
  (2020) 129}, \href{http://arxiv.org/abs/2005.13433}{{\ttfamily
  arXiv:2005.13433 [hep-th]}}.

\bibitem{deBoer:2003vf}
J.~de~Boer and S.~N. Solodukhin, ``{A Holographic reduction of Minkowski
  space-time},'' \href{http://dx.doi.org/10.1016/S0550-3213(03)00494-2}{{\em
  Nucl. Phys.} {\bfseries B665} (2003) 545--593},
\href{http://arxiv.org/abs/hep-th/0303006}{{\ttfamily arXiv:hep-th/0303006
  [hep-th]}}.

\bibitem{Cheung:2016iub}
C.~Cheung, A.~de~la Fuente, and R.~Sundrum, ``{4D scattering amplitudes and
  asymptotic symmetries from 2D CFT},''
  \href{http://dx.doi.org/10.1007/JHEP01(2017)112}{{\em JHEP} {\bfseries 01}
  (2017) 112},
\href{http://arxiv.org/abs/1609.00732}{{\ttfamily arXiv:1609.00732 [hep-th]}}.

\bibitem{Pasterski:2017ylz}
S.~Pasterski, S.-H. Shao, and A.~Strominger, ``{Gluon Amplitudes as 2d
  Conformal Correlators},''
  \href{http://dx.doi.org/10.1103/PhysRevD.96.085006}{{\em Phys. Rev.}
  {\bfseries D96} no.~8, (2017) 085006},
\href{http://arxiv.org/abs/1706.03917}{{\ttfamily arXiv:1706.03917 [hep-th]}}.

\bibitem{ss}
S.~Pasterski, ``{Soft Shadows},'' {\em
  \href{https://physicsgirl.com/ss.pdf}{978-0-9863685-4-7}} (2017) .

\bibitem{Fan:2021isc}
W.~Fan, A.~Fotopoulos, S.~Stieberger, T.~R. Taylor, and B.~Zhu, ``{Conformal
  Blocks from Celestial Gluon Amplitudes},''
  \href{http://arxiv.org/abs/2103.04420}{{\ttfamily arXiv:2103.04420
  [hep-th]}}.

\bibitem{Atanasov:2021cje}
A.~Atanasov, W.~Melton, A.-M. Raclariu, and A.~Strominger, ``{Conformal Block
  Expansion in Celestial CFT},''
  \href{http://arxiv.org/abs/2104.13432}{{\ttfamily arXiv:2104.13432
  [hep-th]}}.

\bibitem{Crawley:2021ivb}
E.~Crawley, N.~Miller, S.~A. Narayanan, and A.~Strominger, ``{State-Operator
  Correspondence in Celestial Conformal Field Theory},''
  \href{http://arxiv.org/abs/2105.00331}{{\ttfamily arXiv:2105.00331
  [hep-th]}}.

\bibitem{Sharma:2021gcz}
A.~Sharma, ``{Ambidextrous light transforms for celestial amplitudes},''
  \href{http://arxiv.org/abs/2107.06250}{{\ttfamily arXiv:2107.06250
  [hep-th]}}.

\bibitem{Kapec:2021eug}
D.~Kapec and P.~Mitra, ``{Shadows and Soft Exchange in Celestial CFT},''
  \href{http://arxiv.org/abs/2109.00073}{{\ttfamily arXiv:2109.00073
  [hep-th]}}.

\bibitem{Fotopoulos:2019tpe}
A.~Fotopoulos and T.~R. Taylor, ``{Primary Fields in Celestial CFT},''
  \href{http://dx.doi.org/10.1007/JHEP10(2019)167}{{\em JHEP} {\bfseries 10}
  (2019) 167},
\href{http://arxiv.org/abs/1906.10149}{{\ttfamily arXiv:1906.10149 [hep-th]}}.

\bibitem{Donnay:2020guq}
L.~Donnay, S.~Pasterski, and A.~Puhm, ``{Asymptotic Symmetries and Celestial
  CFT},'' \href{http://dx.doi.org/10.1007/JHEP09(2020)176}{{\em JHEP}
  {\bfseries 09} (2020) 176}, \href{http://arxiv.org/abs/2005.08990}{{\ttfamily
  arXiv:2005.08990 [hep-th]}}.

\bibitem{Donnay:2018neh}
L.~Donnay, A.~Puhm, and A.~Strominger, ``{Conformally Soft Photons and
  Gravitons},'' \href{http://dx.doi.org/10.1007/JHEP01(2019)184}{{\em JHEP}
  {\bfseries 01} (2019) 184},
\href{http://arxiv.org/abs/1810.05219}{{\ttfamily arXiv:1810.05219 [hep-th]}}.

\bibitem{Stieberger:2018onx}
S.~Stieberger and T.~R. Taylor, ``{Symmetries of Celestial Amplitudes},''
  \href{http://dx.doi.org/10.1016/j.physletb.2019.03.063}{{\em Phys. Lett. B}
  {\bfseries 793} (2019) 141--143},
  \href{http://arxiv.org/abs/1812.01080}{{\ttfamily arXiv:1812.01080
  [hep-th]}}.

\bibitem{Law:2019glh}
Y.~T.~A. Law and M.~Zlotnikov, ``{Poincar\'e Constraints on Celestial
  Amplitudes},''
\href{http://arxiv.org/abs/1910.04356}{{\ttfamily arXiv:1910.04356 [hep-th]}}.

\bibitem{Adamo:2019ipt}
T.~Adamo, L.~Mason, and A.~Sharma, ``{Celestial amplitudes and conformal soft
  theorems},'' \href{http://dx.doi.org/10.1088/1361-6382/ab42ce}{{\em Class.
  Quant. Grav.} {\bfseries 36} no.~20, (2019) 205018},
  \href{http://arxiv.org/abs/1905.09224}{{\ttfamily arXiv:1905.09224
  [hep-th]}}.

\bibitem{Ball:2019atb}
A.~Ball, E.~Himwich, S.~A. Narayanan, S.~Pasterski, and A.~Strominger,
  ``{Uplifting AdS$_{3}$/CFT$_{2}$ to flat space holography},''
  \href{http://dx.doi.org/10.1007/JHEP08(2019)168}{{\em JHEP} {\bfseries 08}
  (2019) 168}, \href{http://arxiv.org/abs/1905.09809}{{\ttfamily
  arXiv:1905.09809 [hep-th]}}.

\bibitem{Guevara:2019ypd}
A.~Guevara, ``{Notes on Conformal Soft Theorems and Recursion Relations in
  Gravity},''
\href{http://arxiv.org/abs/1906.07810}{{\ttfamily arXiv:1906.07810 [hep-th]}}.

\bibitem{Pate:2019lpp}
M.~Pate, A.-M. Raclariu, A.~Strominger, and E.~Y. Yuan, ``{Celestial Operator
  Products of Gluons and Gravitons},''
  \href{http://arxiv.org/abs/1910.07424}{{\ttfamily arXiv:1910.07424
  [hep-th]}}.

\bibitem{Fotopoulos:2020bqj}
A.~Fotopoulos, S.~Stieberger, T.~R. Taylor, and B.~Zhu, ``{Extended Super BMS
  Algebra of Celestial CFT},''
  \href{http://dx.doi.org/10.1007/JHEP09(2020)198}{{\em JHEP} {\bfseries 09}
  (2020) 198}, \href{http://arxiv.org/abs/2007.03785}{{\ttfamily
  arXiv:2007.03785 [hep-th]}}.

\bibitem{Guevara:2021abz}
A.~Guevara, E.~Himwich, M.~Pate, and A.~Strominger, ``{Holographic Symmetry
  Algebras for Gauge Theory and Gravity},''
  \href{http://arxiv.org/abs/2103.03961}{{\ttfamily arXiv:2103.03961
  [hep-th]}}.

\bibitem{Banerjee:2018gce}
S.~Banerjee, ``{Null Infinity and Unitary Representation of The Poincare
  Group},'' \href{http://dx.doi.org/10.1007/JHEP01(2019)205}{{\em JHEP}
  {\bfseries 01} (2019) 205}, \href{http://arxiv.org/abs/1801.10171}{{\ttfamily
  arXiv:1801.10171 [hep-th]}}.

\bibitem{Banerjee:2018fgd}
S.~Banerjee, ``{Symmetries of free massless particles and soft theorems},''
  \href{http://dx.doi.org/10.1007/s10714-019-2609-z}{{\em Gen. Rel. Grav.}
  {\bfseries 51} no.~9, (2019) 128},
  \href{http://arxiv.org/abs/1804.06646}{{\ttfamily arXiv:1804.06646
  [hep-th]}}.

\bibitem{Pasterski:2021fjn}
S.~Pasterski, A.~Puhm, and E.~Trevisani, ``{Celestial Diamonds: Conformal
  Multiplets in Celestial CFT},''
  \href{http://arxiv.org/abs/2105.03516}{{\ttfamily arXiv:2105.03516
  [hep-th]}}.

\bibitem{Pasterski:2021dqe}
S.~Pasterski, A.~Puhm, and E.~Trevisani, ``{Revisiting the Conformally Soft
  Sector with Celestial Diamonds},''
  \href{http://arxiv.org/abs/2105.09792}{{\ttfamily arXiv:2105.09792
  [hep-th]}}.

\bibitem{Fan:2019emx}
W.~Fan, A.~Fotopoulos, and T.~R. Taylor, ``{Soft Limits of Yang-Mills
  Amplitudes and Conformal Correlators},''
  \href{http://dx.doi.org/10.1007/JHEP05(2019)121}{{\em JHEP} {\bfseries 05}
  (2019) 121}, \href{http://arxiv.org/abs/1903.01676}{{\ttfamily
  arXiv:1903.01676 [hep-th]}}.

\bibitem{Pate:2019mfs}
M.~Pate, A.-M. Raclariu, and A.~Strominger, ``{Conformally Soft Theorem in
  Gauge Theory},'' \href{http://dx.doi.org/10.1103/PhysRevD.100.085017}{{\em
  Phys. Rev.} {\bfseries D100} no.~8, (2019) 085017},
\href{http://arxiv.org/abs/1904.10831}{{\ttfamily arXiv:1904.10831 [hep-th]}}.

\bibitem{Puhm:2019zbl}
A.~Puhm, ``{Conformally Soft Theorem in Gravity},''
  \href{http://dx.doi.org/10.1007/JHEP09(2020)130}{{\em JHEP} {\bfseries 09}
  (2020) 130}, \href{http://arxiv.org/abs/1905.09799}{{\ttfamily
  arXiv:1905.09799 [hep-th]}}.

\bibitem{Stieberger:2018edy}
S.~Stieberger and T.~R. Taylor, ``{Strings on Celestial Sphere},''
  \href{http://dx.doi.org/10.1016/j.nuclphysb.2018.08.019}{{\em Nucl. Phys.}
  {\bfseries B935} (2018) 388--411},
\href{http://arxiv.org/abs/1806.05688}{{\ttfamily arXiv:1806.05688 [hep-th]}}.

\bibitem{Arkani-Hamed:2020gyp}
N.~Arkani-Hamed, M.~Pate, A.-M. Raclariu, and A.~Strominger, ``{Celestial
  Amplitudes from UV to IR},''
  \href{http://arxiv.org/abs/2012.04208}{{\ttfamily arXiv:2012.04208
  [hep-th]}}.

\bibitem{Chang:2021wvv}
C.-M. Chang, Y.-t. Huang, Z.-X. Huang, and W.~Li, ``{Bulk locality from the
  celestial amplitude},'' \href{http://arxiv.org/abs/2106.11948}{{\ttfamily
  arXiv:2106.11948 [hep-th]}}.

\bibitem{Fotopoulos:2019vac}
A.~Fotopoulos, S.~Stieberger, T.~R. Taylor, and B.~Zhu, ``{Extended BMS Algebra
  of Celestial CFT},'' \href{http://dx.doi.org/10.1007/JHEP03(2020)130}{{\em
  JHEP} {\bfseries 03} (2020) 130},
  \href{http://arxiv.org/abs/1912.10973}{{\ttfamily arXiv:1912.10973
  [hep-th]}}.

\bibitem{Banerjee:2020kaa}
S.~Banerjee, S.~Ghosh, and R.~Gonzo, ``{BMS symmetry of celestial OPE},''
  \href{http://dx.doi.org/10.1007/JHEP04(2020)130}{{\em JHEP} {\bfseries 04}
  (2020) 130}, \href{http://arxiv.org/abs/2002.00975}{{\ttfamily
  arXiv:2002.00975 [hep-th]}}.

\bibitem{Fan:2020xjj}
W.~Fan, A.~Fotopoulos, S.~Stieberger, and T.~R. Taylor, ``{On Sugawara
  construction on Celestial Sphere},''
  \href{http://dx.doi.org/10.1007/JHEP09(2020)139}{{\em JHEP} {\bfseries 09}
  (2020) 139}, \href{http://arxiv.org/abs/2005.10666}{{\ttfamily
  arXiv:2005.10666 [hep-th]}}.

\bibitem{Ebert:2020nqf}
S.~Ebert, A.~Sharma, and D.~Wang, ``{Descendants in celestial CFT and emergent
  multi-collinear factorization},''
  \href{http://arxiv.org/abs/2009.07881}{{\ttfamily arXiv:2009.07881
  [hep-th]}}.

\bibitem{Himwich:2021dau}
E.~Himwich, M.~Pate, and K.~Singh, ``{Celestial Operator Product Expansions and
  ${\rm w}_{1+\infty}$ Symmetry for All Spins},''
  \href{http://arxiv.org/abs/2108.07763}{{\ttfamily arXiv:2108.07763
  [hep-th]}}.

\bibitem{Guevara:2021tvr}
A.~Guevara, ``{Celestial OPE blocks},''
  \href{http://arxiv.org/abs/2108.12706}{{\ttfamily arXiv:2108.12706
  [hep-th]}}.

\bibitem{Banerjee:2020zlg}
S.~Banerjee, S.~Ghosh, and P.~Paul, ``{MHV Graviton Scattering Amplitudes and
  Current Algebra on the Celestial Sphere},''
  \href{http://arxiv.org/abs/2008.04330}{{\ttfamily arXiv:2008.04330
  [hep-th]}}.

\bibitem{Banerjee:2020vnt}
S.~Banerjee and S.~Ghosh, ``{MHV Gluon Scattering Amplitudes from Celestial
  Current Algebras},'' \href{http://arxiv.org/abs/2011.00017}{{\ttfamily
  arXiv:2011.00017 [hep-th]}}.

\bibitem{Banerjee:2021dlm}
S.~Banerjee, S.~Ghosh, and P.~Paul, ``{(Chiral) Virasoro invariance of the
  tree-level MHV graviton scattering amplitudes},''
  \href{http://arxiv.org/abs/2108.04262}{{\ttfamily arXiv:2108.04262
  [hep-th]}}.

\bibitem{Freidel:2021qpz}
L.~Freidel and D.~Pranzetti, ``{Gravity from symmetry: Duality and impulsive
  waves},'' \href{http://arxiv.org/abs/2109.06342}{{\ttfamily arXiv:2109.06342
  [hep-th]}}.

\bibitem{Banerjee:2017jeg}
N.~Banerjee, S.~Banerjee, S.~Atul~Bhatkar, and S.~Jain, ``{Conformal Structure
  of Massless Scalar Amplitudes Beyond Tree level},''
  \href{http://dx.doi.org/10.1007/JHEP04(2018)039}{{\em JHEP} {\bfseries 04}
  (2018) 039},
\href{http://arxiv.org/abs/1711.06690}{{\ttfamily arXiv:1711.06690 [hep-th]}}.

\bibitem{Gonzalez:2020tpi}
H.~A. Gonz\'alez, A.~Puhm, and F.~Rojas, ``{Loop corrections to celestial
  amplitudes},'' \href{http://dx.doi.org/10.1103/PhysRevD.102.126027}{{\em
  Phys. Rev. D} {\bfseries 102} no.~12, (2020) 126027},
  \href{http://arxiv.org/abs/2009.07290}{{\ttfamily arXiv:2009.07290
  [hep-th]}}.

\bibitem{Albayrak:2020saa}
S.~Albayrak, C.~Chowdhury, and S.~Kharel, ``{On loop celestial amplitudes for
  gauge theory and gravity},''
  \href{http://dx.doi.org/10.1103/PhysRevD.102.126020}{{\em Phys. Rev. D}
  {\bfseries 102} (2020) 126020},
  \href{http://arxiv.org/abs/2007.09338}{{\ttfamily arXiv:2007.09338
  [hep-th]}}.

\bibitem{Casali:2020vuy}
E.~Casali and A.~Puhm, ``{A Double Copy for Celestial Amplitudes},''
  \href{http://arxiv.org/abs/2007.15027}{{\ttfamily arXiv:2007.15027
  [hep-th]}}.

\bibitem{Casali:2020uvr}
E.~Casali and A.~Sharma, ``{Celestial double copy from the worldsheet},''
  \href{http://arxiv.org/abs/2011.10052}{{\ttfamily arXiv:2011.10052
  [hep-th]}}.

\bibitem{Kalyanapuram:2020epb}
N.~Kalyanapuram, ``{Soft Gravity by Squaring Soft QED on the Celestial
  Sphere},'' \href{http://dx.doi.org/10.1103/PhysRevD.103.085016}{{\em Phys.
  Rev. D} {\bfseries 103} no.~8, (2021) 085016},
  \href{http://arxiv.org/abs/2011.11412}{{\ttfamily arXiv:2011.11412
  [hep-th]}}.

\bibitem{Pasterski:2020pdk}
S.~Pasterski and A.~Puhm, ``{Shifting Spin on the Celestial Sphere},''
  \href{http://arxiv.org/abs/2012.15694}{{\ttfamily arXiv:2012.15694
  [hep-th]}}.

\bibitem{Brandhuber:2021nez}
A.~Brandhuber, G.~R. Brown, J.~Gowdy, B.~Spence, and G.~Travaglini,
  ``{Celestial Superamplitudes},''
  \href{http://arxiv.org/abs/2105.10263}{{\ttfamily arXiv:2105.10263
  [hep-th]}}.

\bibitem{Jiang:2021xzy}
H.~Jiang, ``{Celestial superamplitude in $\mathcal N=4$ SYM theory},''
  \href{http://arxiv.org/abs/2105.10269}{{\ttfamily arXiv:2105.10269
  [hep-th]}}.

\bibitem{Hu:2021lrx}
Y.~Hu, L.~Ren, A.~Y. Srikant, and A.~Volovich, ``{Celestial Dual Superconformal
  Symmetry, MHV Amplitudes and Differential Equations},''
  \href{http://arxiv.org/abs/2106.16111}{{\ttfamily arXiv:2106.16111
  [hep-th]}}.

\bibitem{Ferro:2021dub}
L.~Ferro and R.~Moerman, ``{The Grassmannian for Celestial Superamplitudes},''
  \href{http://arxiv.org/abs/2107.07496}{{\ttfamily arXiv:2107.07496
  [hep-th]}}.

\bibitem{Jiang:2021ovh}
H.~Jiang, ``{Holographic Chiral Algebra: Supersymmetry, Infinite Ward
  Identities, and EFTs},'' \href{http://arxiv.org/abs/2108.08799}{{\ttfamily
  arXiv:2108.08799 [hep-th]}}.

\bibitem{Pano:2021ewd}
Y.~Pano, S.~Pasterski, and A.~Puhm, ``{Conformally Soft Fermions},''
  \href{http://arxiv.org/abs/2108.11422}{{\ttfamily arXiv:2108.11422
  [hep-th]}}.

\bibitem{Atanasov:2021oyu}
A.~Atanasov, A.~Ball, W.~Melton, A.-M. Raclariu, and A.~Strominger, ``{(2, 2)
  Scattering and the celestial torus},''
  \href{http://dx.doi.org/10.1007/JHEP07(2021)083}{{\em JHEP} {\bfseries 07}
  (2021) 083}, \href{http://arxiv.org/abs/2101.09591}{{\ttfamily
  arXiv:2101.09591 [hep-th]}}.

\bibitem{Bhattacharjee:2021mdc}
B.~Bhattacharjee and C.~Krishnan, ``{Celestial Klein Spaces},''
  \href{http://arxiv.org/abs/2110.06180}{{\ttfamily arXiv:2110.06180
  [hep-th]}}.

\bibitem{Adamo:2021lrv}
T.~Adamo, L.~Mason, and A.~Sharma, ``{Celestial $w_{1+\infty}$ symmetries from
  twistor space},'' \href{http://arxiv.org/abs/2110.06066}{{\ttfamily
  arXiv:2110.06066 [hep-th]}}.

\bibitem{Adamo:2021zpw}
T.~Adamo, W.~Bu, E.~Casali, and A.~Sharma, ``{Celestial operator products from
  the worldsheet},'' \href{http://arxiv.org/abs/2111.02279}{{\ttfamily
  arXiv:2111.02279 [hep-th]}}.

\bibitem{Schreiber:2017jsr}
A.~Schreiber, A.~Volovich, and M.~Zlotnikov, ``{Tree-level gluon amplitudes on
  the celestial sphere},''
  \href{http://dx.doi.org/10.1016/j.physletb.2018.04.010}{{\em Phys. Lett.}
  {\bfseries B781} (2018) 349--357},
\href{http://arxiv.org/abs/1711.08435}{{\ttfamily arXiv:1711.08435 [hep-th]}}.

\bibitem{Zlotnikov2018}
M.~Zlotnikov, \href{http://dx.doi.org/10.26300/a7hm-2x50}{{\em {Scattering
  Equations, Soft Theorems, and Amplitudes on the Celestial Sphere}}}.
\newblock PhD thesis, Brown U., 2018.

\bibitem{Melton:2021kkz}
W.~Melton, ``{Celestial Feynman Rules for Scalars},''
  \href{http://arxiv.org/abs/2109.07462}{{\ttfamily arXiv:2109.07462
  [hep-th]}}.

\bibitem{Lam:2017ofc}
H.~T. Lam and S.-H. Shao, ``{Conformal Basis, Optical Theorem, and the Bulk
  Point Singularity},''
\href{http://arxiv.org/abs/1711.06138}{{\ttfamily arXiv:1711.06138 [hep-th]}}.

\bibitem{Law:2020xcf}
Y.~A. Law and M.~Zlotnikov, ``{Relativistic partial waves for celestial
  amplitudes},'' \href{http://dx.doi.org/10.1007/JHEP11(2020)149}{{\em JHEP}
  {\bfseries 11} (2020) 149}, \href{http://arxiv.org/abs/2008.02331}{{\ttfamily
  arXiv:2008.02331 [hep-th]}}.

\bibitem{Banerjee:2021cly}
S.~Banerjee, S.~Ghosh, and S.~Satyam~Samal, ``{Subsubleading soft graviton
  symmetry and MHV graviton scattering amplitudes},''
  \href{http://arxiv.org/abs/2104.02546}{{\ttfamily arXiv:2104.02546
  [hep-th]}}.

\bibitem{Strominger:2021lvk}
A.~Strominger, ``{w(1+infinity) and the Celestial Sphere},''
  \href{http://arxiv.org/abs/2105.14346}{{\ttfamily arXiv:2105.14346
  [hep-th]}}.

\bibitem{Gupta:2021cwo}
N.~Gupta, P.~Paul, and N.~V. Suryanarayana, ``{An $\widehat{sl_2}$ Symmetry of
  ${\mathbb R}^{1,3}$ Gravity},''
  \href{http://arxiv.org/abs/2109.06857}{{\ttfamily arXiv:2109.06857
  [hep-th]}}.

\bibitem{Jiang:2021csc}
H.~Jiang, ``{Celestial OPEs and $ w_{1+\infty}$ algebra from worldsheet in
  string theory},'' \href{http://arxiv.org/abs/2110.04255}{{\ttfamily
  arXiv:2110.04255 [hep-th]}}.

\bibitem{Banerjee:2021uxe}
N.~Banerjee, T.~Rahnuma, and R.~K. Singh, ``{Asymptotic Symmetry of Four
  Dimensional Einstein-Yang-Mills and Einstein-Maxwell Theory},''
  \href{http://arxiv.org/abs/2110.15657}{{\ttfamily arXiv:2110.15657
  [hep-th]}}.

\bibitem{Ahn:2021erj}
C.~Ahn, ``{Towards A Supersymmetric $w_{1+\infty}$ Symmetry in the Celestial
  Conformal Field Theory},'' \href{http://arxiv.org/abs/2111.04268}{{\ttfamily
  arXiv:2111.04268 [hep-th]}}.

\bibitem{Banerjee:2019prz}
S.~Banerjee, S.~Ghosh, P.~Pandey, and A.~P. Saha, ``{Modified celestial
  amplitude in Einstein gravity},''
  \href{http://dx.doi.org/10.1007/JHEP03(2020)125}{{\em JHEP} {\bfseries 03}
  (2020) 125}, \href{http://arxiv.org/abs/1909.03075}{{\ttfamily
  arXiv:1909.03075 [hep-th]}}.

\bibitem{Liu:2021tif}
C.~Liu and D.~A. Lowe, ``{Conformal Wave Expansions for Flat Space
  Amplitudes},'' \href{http://arxiv.org/abs/2105.01026}{{\ttfamily
  arXiv:2105.01026 [hep-th]}}.

\bibitem{Lippstreu:2021avq}
L.~Lippstreu, ``{Zwanziger's pairwise little group on the celestial sphere},''
  \href{http://arxiv.org/abs/2106.00084}{{\ttfamily arXiv:2106.00084
  [hep-th]}}.

\bibitem{Liu:2021xlm}
C.~Liu and D.~A. Lowe, ``{Conformal Wavefunctions for Graviton Amplitudes},''
  \href{http://arxiv.org/abs/2109.00037}{{\ttfamily arXiv:2109.00037
  [hep-th]}}.

\bibitem{Nguyen:2020hot}
K.~Nguyen and J.~Salzer, ``{The effective action of superrotation modes},''
  \href{http://dx.doi.org/10.1007/JHEP02(2021)108}{{\em JHEP} {\bfseries 02}
  (2021) 108}, \href{http://arxiv.org/abs/2008.03321}{{\ttfamily
  arXiv:2008.03321 [hep-th]}}.

\bibitem{Kalyanapuram:2020aya}
N.~Kalyanapuram, ``{Gauge and Gravity Amplitudes on the Celestial Sphere},''
  \href{http://arxiv.org/abs/2012.04579}{{\ttfamily arXiv:2012.04579
  [hep-th]}}.

\bibitem{Magnea:2021fvy}
L.~Magnea, ``{Non-abelian infrared divergences on the celestial sphere},''
  \href{http://dx.doi.org/10.1007/JHEP05(2021)282}{{\em JHEP} {\bfseries 05}
  (2021) 282}, \href{http://arxiv.org/abs/2104.10254}{{\ttfamily
  arXiv:2104.10254 [hep-th]}}.

\bibitem{Gonzalez:2021dxw}
H.~A. Gonz\'alez and F.~Rojas, ``{The structure of IR divergences in celestial
  gluon amplitudes},'' \href{http://dx.doi.org/10.1007/JHEP06(2021)171}{{\em
  JHEP} {\bfseries 2021} no.~06, (2021) 171},
  \href{http://arxiv.org/abs/2104.12979}{{\ttfamily arXiv:2104.12979
  [hep-th]}}.

\bibitem{Kalyanapuram:2021bvf}
N.~Kalyanapuram, ``{Holographic soft S-matrix in QED and gravity},''
  \href{http://dx.doi.org/10.1103/PhysRevD.104.045006}{{\em Phys. Rev. D}
  {\bfseries 104} no.~4, (2021) 045006},
  \href{http://arxiv.org/abs/2105.04314}{{\ttfamily arXiv:2105.04314
  [hep-th]}}.

\bibitem{Nguyen:2021qkt}
K.~Nguyen and J.~Salzer, ``{Celestial IR divergences and the effective action
  of supertranslation modes},''
  \href{http://arxiv.org/abs/2105.10526}{{\ttfamily arXiv:2105.10526
  [hep-th]}}.

\bibitem{Kalyanapuram:2021tnl}
N.~Kalyanapuram, ``{Infrared and Holographic Aspects of the $S$-Matrix in Gauge
  Theory and Gravity},'' \href{http://arxiv.org/abs/2107.06660}{{\ttfamily
  arXiv:2107.06660 [hep-th]}}.

\bibitem{Kalyanapuram:2021fic}
N.~Kalyanapuram, ``{Holographic Representations of Supertranslation
  Eigenstates},'' \href{http://arxiv.org/abs/2109.12351}{{\ttfamily
  arXiv:2109.12351 [hep-th]}}.

\bibitem{Kalyanapuram:2021dql}
N.~Kalyanapuram, ``{Spin Multipole Expansion of the Memory Effect},''
  \href{http://arxiv.org/abs/2109.12368}{{\ttfamily arXiv:2109.12368
  [hep-th]}}.

\bibitem{Nastase:2021izh}
H.~Nastase, F.~Rojas, and C.~Rubio, ``{Celestial IR divergences in general
  most-subleading-color gluon and gravity amplitudes},''
  \href{http://arxiv.org/abs/2111.06861}{{\ttfamily arXiv:2111.06861
  [hep-th]}}.

\bibitem{Nandan:2019jas}
D.~Nandan, A.~Schreiber, A.~Volovich, and M.~Zlotnikov, ``{Celestial
  Amplitudes: Conformal Partial Waves and Soft Limits},''
  \href{http://dx.doi.org/10.1007/JHEP10(2019)018}{{\em JHEP} {\bfseries 10}
  (2019) 018},
\href{http://arxiv.org/abs/1904.10940}{{\ttfamily arXiv:1904.10940 [hep-th]}}.

\bibitem{Fan:2021pbp}
W.~Fan, A.~Fotopoulos, S.~Stieberger, T.~R. Taylor, and B.~Zhu, ``{Conformal
  Blocks from Celestial Gluon Amplitudes II: Single-valued Correlators},''
  \href{http://arxiv.org/abs/2108.10337}{{\ttfamily arXiv:2108.10337
  [hep-th]}}.

\bibitem{Hubner:2019sly}
M.~H\"ubner, C.~Talbot, P.~D. Lasky, and E.~Thrane, ``{Measuring
  gravitational-wave memory in the first LIGO/Virgo gravitational-wave
  transient catalog},''
  \href{http://dx.doi.org/10.1103/PhysRevD.101.023011}{{\em Phys. Rev. D}
  {\bfseries 101} no.~2, (2020) 023011},
  \href{http://arxiv.org/abs/1911.12496}{{\ttfamily arXiv:1911.12496
  [astro-ph.HE]}}.

\bibitem{Mitman:2021xkq}
K.~Mitman {\em et~al.}, ``{Fixing the BMS frame of numerical relativity
  waveforms},'' \href{http://dx.doi.org/10.1103/PhysRevD.104.024051}{{\em Phys.
  Rev. D} {\bfseries 104} no.~2, (2021) 024051},
  \href{http://arxiv.org/abs/2105.02300}{{\ttfamily arXiv:2105.02300 [gr-qc]}}.

\bibitem{Accardi:2012qut}
A.~Accardi {\em et~al.}, ``{Electron Ion Collider: The Next QCD Frontier}:
  {Understanding the glue that binds us all},''
  \href{http://dx.doi.org/10.1140/epja/i2016-16268-9}{{\em Eur. Phys. J. A}
  {\bfseries 52} no.~9, (2016) 268},
  \href{http://arxiv.org/abs/1212.1701}{{\ttfamily arXiv:1212.1701 [nucl-ex]}}.

\bibitem{AbdulKhalek:2021gbh}
R.~Abdul~Khalek {\em et~al.}, ``{Science Requirements and Detector Concepts for
  the Electron-Ion Collider: EIC Yellow Report},''
  \href{http://arxiv.org/abs/2103.05419}{{\ttfamily arXiv:2103.05419
  [physics.ins-det]}}.

\bibitem{Kapec:2017tkm}
D.~Kapec, M.~Perry, A.-M. Raclariu, and A.~Strominger, ``{Infrared Divergences
  in QED, Revisited},''
  \href{http://dx.doi.org/10.1103/PhysRevD.96.085002}{{\em Phys. Rev. D}
  {\bfseries 96} no.~8, (2017) 085002},
  \href{http://arxiv.org/abs/1705.04311}{{\ttfamily arXiv:1705.04311
  [hep-th]}}.

\bibitem{Chung:1965zza}
V.~Chung, ``{Infrared Divergence in Quantum Electrodynamics},''
  \href{http://dx.doi.org/10.1103/PhysRev.140.B1110}{{\em Phys. Rev.}
  {\bfseries 140} (1965) B1110--B1122}.

\bibitem{Kulish:1970ut}
P.~P. Kulish and L.~D. Faddeev, ``{Asymptotic conditions and infrared
  divergences in quantum electrodynamics},''
  \href{http://dx.doi.org/10.1007/BF01066485}{{\em Theor. Math. Phys.}
  {\bfseries 4} (1970) 745}.

\bibitem{Gomez:2016hxz}
C.~Gomez and M.~Panchenko, ``{Asymptotic dynamics, large gauge transformations
  and infrared symmetries},'' \href{http://arxiv.org/abs/1608.05630}{{\ttfamily
  arXiv:1608.05630 [hep-th]}}.

\bibitem{Choi:2017bna}
S.~Choi, U.~Kol, and R.~Akhoury, ``{Asymptotic Dynamics in Perturbative Quantum
  Gravity and BMS Supertranslations},''
  \href{http://dx.doi.org/10.1007/JHEP01(2018)142}{{\em JHEP} {\bfseries 01}
  (2018) 142}, \href{http://arxiv.org/abs/1708.05717}{{\ttfamily
  arXiv:1708.05717 [hep-th]}}.

\bibitem{Choi:2017ylo}
S.~Choi and R.~Akhoury, ``{BMS Supertranslation Symmetry Implies Faddeev-Kulish
  Amplitudes},'' \href{http://dx.doi.org/10.1007/JHEP02(2018)171}{{\em JHEP}
  {\bfseries 02} (2018) 171}, \href{http://arxiv.org/abs/1712.04551}{{\ttfamily
  arXiv:1712.04551 [hep-th]}}.

\bibitem{Gomez:2018war}
C.~G\'omez, R.~Letschka, and S.~Zell, ``{The Scales of the Infrared},''
  \href{http://dx.doi.org/10.1007/JHEP09(2018)115}{{\em JHEP} {\bfseries 09}
  (2018) 115}, \href{http://arxiv.org/abs/1807.07079}{{\ttfamily
  arXiv:1807.07079 [hep-th]}}.

\bibitem{Choi:2019rlz}
S.~Choi and R.~Akhoury, ``{Subleading soft dressings of asymptotic states in
  QED and perturbative quantum gravity},''
  \href{http://dx.doi.org/10.1007/JHEP09(2019)031}{{\em JHEP} {\bfseries 09}
  (2019) 031}, \href{http://arxiv.org/abs/1907.05438}{{\ttfamily
  arXiv:1907.05438 [hep-th]}}.

\bibitem{Hawking:2015qqa}
S.~Hawking, ``{The Information Paradox for Black Holes},''
  \href{http://arxiv.org/abs/1509.01147}{{\ttfamily arXiv:1509.01147
  [hep-th]}}.

\bibitem{Hawking:2016msc}
S.~W. Hawking, M.~J. Perry, and A.~Strominger, ``{Soft Hair on Black Holes},''
  \href{http://dx.doi.org/10.1103/PhysRevLett.116.231301}{{\em Phys. Rev.
  Lett.} {\bfseries 116} no.~23, (2016) 231301},
\href{http://arxiv.org/abs/1601.00921}{{\ttfamily arXiv:1601.00921 [hep-th]}}.

\bibitem{Hawking:2016sgy}
S.~W. Hawking, M.~J. Perry, and A.~Strominger, ``{Superrotation Charge and
  Supertranslation Hair on Black Holes},''
  \href{http://dx.doi.org/10.1007/JHEP05(2017)161}{{\em JHEP} {\bfseries 05}
  (2017) 161}, \href{http://arxiv.org/abs/1611.09175}{{\ttfamily
  arXiv:1611.09175 [hep-th]}}.

\bibitem{Strominger:2017aeh}
A.~Strominger, ``{Black Hole Information Revisited},''
  \href{http://arxiv.org/abs/1706.07143}{{\ttfamily arXiv:1706.07143
  [hep-th]}}.

\bibitem{Donnay:2015abr}
L.~Donnay, G.~Giribet, H.~A. Gonzalez, and M.~Pino, ``{Supertranslations and
  Superrotations at the Black Hole Horizon},''
  \href{http://dx.doi.org/10.1103/PhysRevLett.116.091101}{{\em Phys. Rev.
  Lett.} {\bfseries 116} no.~9, (2016) 091101},
\href{http://arxiv.org/abs/1511.08687}{{\ttfamily arXiv:1511.08687 [hep-th]}}.

\bibitem{Mirbabayi:2016axw}
M.~Mirbabayi and M.~Porrati, ``{Dressed Hard States and Black Hole Soft
  Hair},'' \href{http://dx.doi.org/10.1103/PhysRevLett.117.211301}{{\em Phys.
  Rev. Lett.} {\bfseries 117} no.~21, (2016) 211301},
  \href{http://arxiv.org/abs/1607.03120}{{\ttfamily arXiv:1607.03120
  [hep-th]}}.

\bibitem{Donnay:2016ejv}
L.~Donnay, G.~Giribet, H.~A. Gonzalez, and M.~Pino, ``{Extended Symmetries at
  the Black Hole Horizon},''
  \href{http://dx.doi.org/10.1007/JHEP09(2016)100}{{\em JHEP} {\bfseries 09}
  (2016) 100},
\href{http://arxiv.org/abs/1607.05703}{{\ttfamily arXiv:1607.05703 [hep-th]}}.

\bibitem{Bousso:2017dny}
R.~Bousso and M.~Porrati, ``{Soft Hair as a Soft Wig},''
  \href{http://dx.doi.org/10.1088/1361-6382/aa8be2}{{\em Class. Quant. Grav.}
  {\bfseries 34} no.~20, (2017) 204001},
  \href{http://arxiv.org/abs/1706.00436}{{\ttfamily arXiv:1706.00436
  [hep-th]}}.

\bibitem{Donnay:2018ckb}
L.~Donnay, G.~Giribet, H.~A. Gonz\'alez, and A.~Puhm, ``{Black hole memory
  effect},'' \href{http://dx.doi.org/10.1103/PhysRevD.98.124016}{{\em Phys.
  Rev. D} {\bfseries 98} no.~12, (2018) 124016},
  \href{http://arxiv.org/abs/1809.07266}{{\ttfamily arXiv:1809.07266
  [hep-th]}}.

\bibitem{Haco:2018ske}
S.~Haco, S.~W. Hawking, M.~J. Perry, and A.~Strominger, ``{Black Hole Entropy
  and Soft Hair},'' \href{http://dx.doi.org/10.1007/JHEP12(2018)098}{{\em JHEP}
  {\bfseries 12} (2018) 098}, \href{http://arxiv.org/abs/1810.01847}{{\ttfamily
  arXiv:1810.01847 [hep-th]}}.

\bibitem{Haco:2019ggi}
S.~Haco, M.~J. Perry, and A.~Strominger, ``{Kerr-Newman Black Hole Entropy and
  Soft Hair},'' \href{http://arxiv.org/abs/1902.02247}{{\ttfamily
  arXiv:1902.02247 [hep-th]}}.

\bibitem{Rahman:2019bmk}
A.~A. Rahman and R.~M. Wald, ``{Black Hole Memory},''
  \href{http://dx.doi.org/10.1103/PhysRevD.101.124010}{{\em Phys. Rev. D}
  {\bfseries 101} no.~12, (2020) 124010},
  \href{http://arxiv.org/abs/1912.12806}{{\ttfamily arXiv:1912.12806 [gr-qc]}}.

\bibitem{Pasterski:2020xvn}
S.~Pasterski and H.~Verlinde, ``{HPS meets AMPS: How Soft Hair Dissolves the
  Firewall},'' \href{http://arxiv.org/abs/2012.03850}{{\ttfamily
  arXiv:2012.03850 [hep-th]}}.

\bibitem{Kapec:2016aqd}
D.~Kapec, A.-M. Raclariu, and A.~Strominger, ``{Area, Entanglement Entropy and
  Supertranslations at Null Infinity},''
  \href{http://dx.doi.org/10.1088/1361-6382/aa7f12}{{\em Class. Quant. Grav.}
  {\bfseries 34} no.~16, (2017) 165007},
  \href{http://arxiv.org/abs/1603.07706}{{\ttfamily arXiv:1603.07706
  [hep-th]}}.

\bibitem{Laddha:2020kvp}
A.~Laddha, S.~G. Prabhu, S.~Raju, and P.~Shrivastava, ``{The Holographic Nature
  of Null Infinity},''
  \href{http://dx.doi.org/10.21468/SciPostPhys.10.2.041}{{\em SciPost Phys.}
  {\bfseries 10} no.~2, (2021) 041},
  \href{http://arxiv.org/abs/2002.02448}{{\ttfamily arXiv:2002.02448
  [hep-th]}}.

\bibitem{Hijano:2019qmi}
E.~Hijano, ``{Flat space physics from AdS/CFT},''
  \href{http://dx.doi.org/10.1007/JHEP07(2019)132}{{\em JHEP} {\bfseries 07}
  (2019) 132}, \href{http://arxiv.org/abs/1905.02729}{{\ttfamily
  arXiv:1905.02729 [hep-th]}}.

\bibitem{Hijano:2020szl}
E.~Hijano and D.~Neuenfeld, ``{Soft photon theorems from CFT Ward identites in
  the flat limit of AdS/CFT},''
  \href{http://dx.doi.org/10.1007/JHEP11(2020)009}{{\em JHEP} {\bfseries 11}
  (2020) 009}, \href{http://arxiv.org/abs/2005.03667}{{\ttfamily
  arXiv:2005.03667 [hep-th]}}.

\bibitem{Compere:2019bua}
G.~Comp\`ere, A.~Fiorucci, and R.~Ruzziconi, ``{The $\Lambda$-BMS$_4$ group of
  dS$_4$ and new boundary conditions for AdS$_4$},''
  \href{http://dx.doi.org/10.1088/1361-6382/ab3d4b}{{\em Class. Quant. Grav.}
  {\bfseries 36} no.~19, (2019) 195017},
  \href{http://arxiv.org/abs/1905.00971}{{\ttfamily arXiv:1905.00971 [gr-qc]}}.
  [Erratum: Class.Quant.Grav. 38, 229501 (2021)].

\end{thebibliography}\endgroup

\end{document}